\documentclass[sn-aps]{sn-jnl}

\usepackage{graphicx}%
\usepackage{multirow}%
\usepackage{amsmath,amssymb,amsfonts}%
\usepackage{amsthm}%
\usepackage{mathrsfs}%
\usepackage[title]{appendix}%
\usepackage{xcolor}%
\usepackage{textcomp}%
\usepackage{manyfoot}%
\usepackage{booktabs}%
\usepackage{algorithm}%
\usepackage{algorithmicx}%
\usepackage{algpseudocode}%
\usepackage{listings}%
\usepackage{subfigure}
\usepackage{bm}

\begin{document}

\title{Recent progress in quantum spin liquids, fractional magnetization plateaus, and unconventional superconductivity in kagome lattices}

\author[1$^{*}$]{\fnm{Li-Wei} \sur{He}}
\email{lwhe@nju.edu.cn}

\author[1,2$^{*}$]{\fnm{Shun-Li} \sur{Yu}}
\email{slyu@nju.edu.cn}

\author[1,2$^{*}$]{\fnm{Jian-Xin} \sur{Li}}
\email{jxli@nju.edu.cn}

\affil[1]{National Laboratory of Solid State Microstructures and Department of Physics, Nanjing University, Nanjing 210093, China}

\affil[2]{Collaborative Innovation Center of Advanced Microstructures, Nanjing University, Nanjing 210093, China}

\abstract{The kagome lattice, with its unique geometric structure, has emerged as a leading platform for exploring quantum many-body physics, particularly in the study of quantum spin liquids (QSLs) and unconventional superconductivity. This review highlights recent advancements in the investigations of QSLs, fractional magnetization plateau phases in kagome antiferromagnets, and unconventional superconductivity in vanadium-based kagome superconductors. We begin by examining the classical ground-state properties of the nearest-neighbor kagome antiferromagnetic Heisenberg model and introducing recent experimental progress in the study of QSLs and fractional magnetization plateau phases. Next, we discuss the fermionic description of the QSL states, along with related gauge theory and the variational Monte Carlo (VMC) method. We then focus on discussing the VMC studies of QSLs and magnetization plateau phases in kagome antiferromagnets. For superconductivity in kagome systems, we first analyze the characteristics of the electronic structure and the possible associated electronic instabilities. Finally, we review recent experimental advances in unconventional superconductivity in AV$_3$Sb$_5$ (A = K, Rb, Cs), with a particular focus on chiral superconductivity and pairing density waves.}


\maketitle

\section{Introduction}\label{sec-intro}

The kagome lattice, characterized by its distinctive two-dimensional arrangement of corner-sharing triangles [see Fig.~\ref{qsl-fig-1}(a)], has  garnered significant interest in condensed matter physics owing to its tendency to host a rich array of novel physical phenomena. This lattice structure serves as an ideal platform for investigating exotic states of matter, particularly quantum spin liquids (QSLs) and unconventional superconductivity, which are beyond traditional understandings of magnetic and superconducting (SC) behaviors.

The $s=1/2$ kagome antiferromagnets are particularly noteworthy, as theoretical models predict the existence of QSL states characterized by long-range entanglement, fractional excitations and the absence of conventional magnetic order even at absolute zero temperature. The concept of QSL originated in the 1970s with Anderson's pioneering resonating valence bond theory laying the theoretical foundation~\cite{Anderson-RVB-1973}. In experimental aspect, the material realization of QSLs on kagome lattices has attracted considerable research interest over the past two decades~\cite{PhysRevLett.98.077204,PhysRevLett.98.107204,Nature.492.7429,science.350.655,Nat.Phys.16.469,PhysRevB.94.060409,PhysRevLett.109.037208,PhysRevB.90.205103,CPL.34.077502,PhysRevB.98.155127,CPL.37.107503,npj.QM.5.74,Nat.Commun.12.3048,PhysRevB.105.L121109,PhysRevB.110.085146,nat.phys.20.1097}. A landmark discovery was the identification of Herbertsmithite, ZnCu$_3$(OH)$_6$Cl$_2$, which exhibits properties consistent with a QSL ground state~\cite{PhysRevLett.98.077204,PhysRevLett.98.107204,Nature.492.7429,science.350.655,Nat.Phys.16.469,PhysRevB.94.060409}.  Furthermore, the potential connections between spin liquids and topological phases of matter have opened new avenues for research. Theoretical investigations have explored the emergence of gauge fields and topological order~\cite{PhysRevB.89.035147,RevModPhys.88.035001,Natl.Sci.Rev.3.68,RevModPhys.89.041004} in these systems, suggesting a rich landscape of quantum phenomena. Thus, the kagome lattice not only serves as a testbed for investigating QSLs but also acts as a platform for exploring fundamental questions about quantum entanglement and topological order. Moreover, when subjected to an external magnetic field, the $s=1/2$ kagome antiferromagnet can also manifest novel quantum states~\cite{PhysRevLett.88.057204,PhysRevB.70.100403,JPSJ.79.053707,PhysRevB.83.100405,nat.com.4.2287,PhysRevB.88.144416,PhysRevB.93.060407,
PhysRevB.98.014415,PhysRevB.98.094423,nat.com.10.1229,PhysRevB.107.L220401,nat.phys.20.435,PhysRevLett.132.226701,PhysRevX.15.021076,pnas.2421390122,PhysRevLett.133.096501,JPSJ.93.123706,arxiv.2503.11834,cpl.42.090704}, further highlighting its potential as an ideal platform for exploring strongly correlated exotic quantum states of matter. Recent research on kagome antiferromagnets has significantly advanced our understanding of their magnetic properties under external magnetic fields, particularly regarding the magnetization plateau phases~\cite{nat.phys.20.435,PhysRevLett.132.226701,PhysRevX.15.021076,pnas.2421390122,PhysRevLett.133.096501,JPSJ.93.123706,arxiv.2503.11834,cpl.42.090704}.

In parallel with QSL investigations, the kagome lattice has also attracted significant interest in the context of superconductivity~\cite{PhysRevB.79.214502,JPCM.23.175702,PhysRevB.85.144402,
PhysRevB.86.121105,PhysRevB.87.115135,PhysRevLett.110.126405,PhysRevLett.125.247002,PhysRevLett.127.187003,Nat.Phys.18.137,CPB.31.097405,nature.612.647,nsr.10.nwac199,nat.rev.phys.5.635,nat.rev.mat.9.420,PhysRevB.84.214527,PhysRevB.86.220510,CPL.39.087401,CPB.33.057401,gui2022lair3ga2}. Its distinctive geometric structure has prompted the exploration of unconventional SC behaviors. Except for the chiral superconductivity realized by doping a Dirac QSL~\cite{PhysRevB.79.214502}, which is born from kagome antiferromegnets, others focus on the superconductivity when the energy band is filled at a van Hove singularity~\cite{PhysRevB.85.144402,PhysRevB.86.121105,PhysRevB.87.115135,PhysRevLett.110.126405}, which related more closely to the peculiar electronic structure of the itinerant kagome system.
Remarkably, the recent discovery of superconductivity in a family of compounds with a kagome net of vanadium atoms AV$_3$Sb$_5$ (A = K, Rb, Cs)~\cite{PhysRevLett.125.247002}, featuring an electronic structure filled to the van Hove singularity,
has sparked a new wave of research into superconductivity~\cite{Nat.Phys.18.137,CPB.31.097405,nature.612.647,nsr.10.nwac199,nat.rev.phys.5.635,nat.rev.mat.9.420,nat.phys.20.534}. Besides the possible unconventional superconductivity, various competing electronic orders in these materials have also been unveiled.


In this brief review, we provide an overview of recent advancements in the investigations of  quantum magnetism and unconventional superconductivity in kagome lattice systems. Regarding quantum magnetism, we highlight recent developments in QSLs and magnetization plateau phases in kagome antiferromagnets. On the topic of unconventional superconductivity, we discuss the progress made in understanding unconventional superconductivity in the AV$_3$Sb$_5$ family of superconductors.

\section{Quantum spin liquids in kagome antiferromagnets and effects of external magnetic field}\label{sec-QSL}

\subsection{$s=1/2$ kagome Heisenberg antiferromagnets: promising candidates of quantum spin liquids}\label{qsl-intro}

The search for QSLs has long been a main focus in condensed matter physics. Unlike magnetically ordered states, which exhibit spontaneous symmetry breaking and local order parameters, QSLs reveal themselves through unique phenomena such as long-range quantum entanglement and fractional spin excitations. These are characterized by emergent particles known as spinons and associated gauge fields. Notably, a gapped QSL is linked to nontrivial topological order. Due to strong geometric frustration and concrete quantum material realizations, the $s=1/2$ kagome Heisenberg antiferromagnets are considered promising candidates for hosting QSLs.

\begin{figure}
    \centering
    \includegraphics[width=1.0\linewidth]{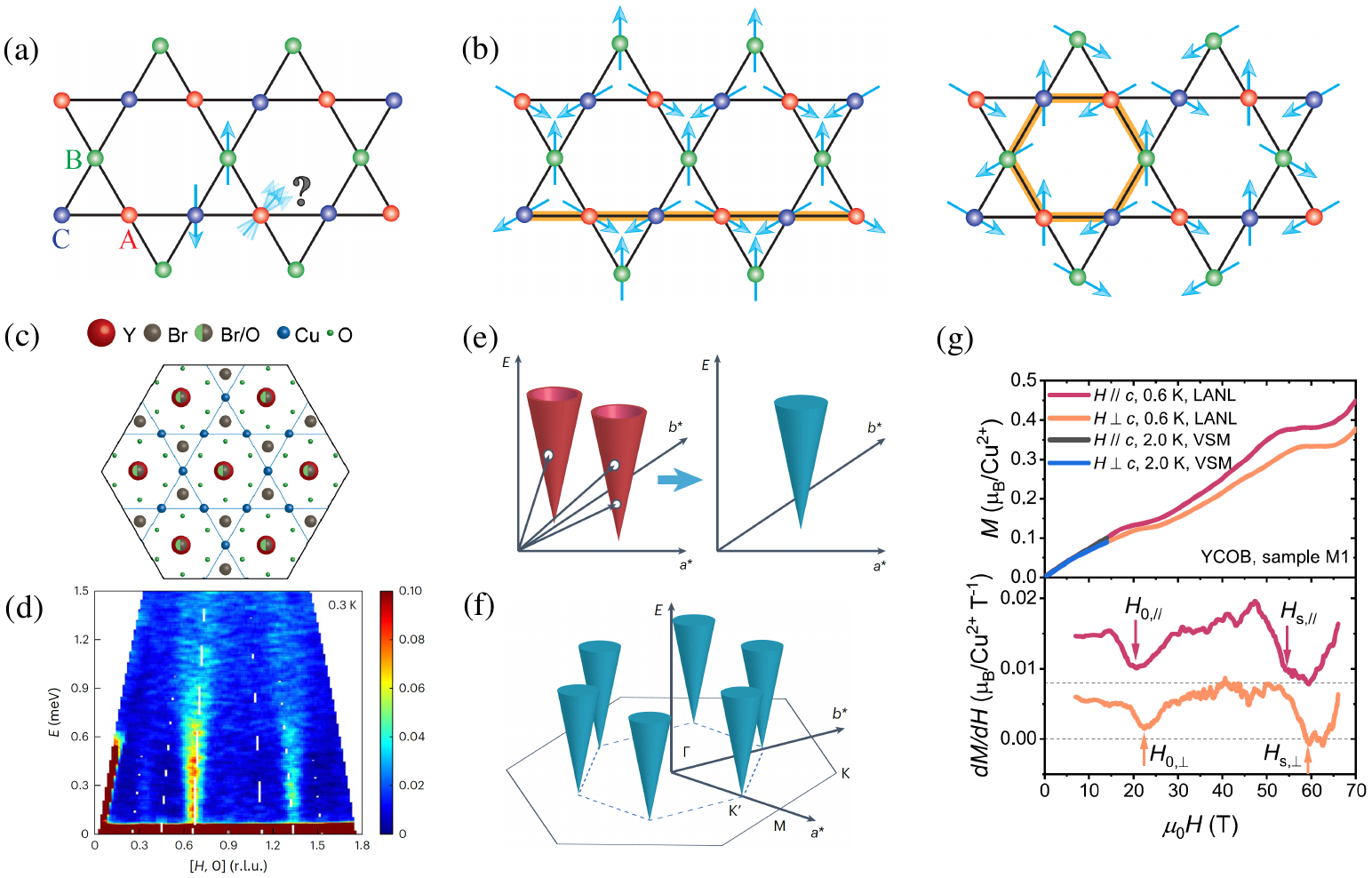}
    \caption{(a) Kagome lattice structure. Each unit cell consists of three sublattices, labeled A, B, and C. For the NN AF Heisenberg model, the system exhibits significant geometric frustration. As illustrated, when two adjacent spins in a unit cell are arranged antiparallel, the third spin encounters a dilemma. (b) Two representative spin configurations with the lowest energy for classical spins. Left: $\boldsymbol{Q} = 0$ state. Right: $\sqrt{3}\times\sqrt{3}$ state. (c) Structure of YCu$_3$(OH)$_6$Br$_2$[Br$_{1-x}$(OH)$_x$]. Cu$^{2+}$ ions form a kagome lattice, and Y$^{3+}$ ions locate at the centers of the stars. Br$^{-}$ are positioned above and below the Y$^{3+}$, randomly mixed with OH$^{-}$. O$^{2-}$ around the hexagons are alternately displaced above and below the kagome plane. Adapted from Ref.~\cite{pnas.2421390122}. (d) Inelastic neutron scattering spectra of YCu$_3$(OH)$_6$Br$_2$[Br$_{1-x}$(OH)$_x$] along the $[H, 0]$ direction at 0.3 K. Adapted from Ref.~\cite{nat.phys.20.1097}. (e) Schematic illustration showing two conical spinon Dirac cones (red) merging into a solid cone spin excitation with a continuum inside (blue). Adapted from Ref.~\cite{nat.phys.20.1097}. (f) Six conical low-energy spin excitations in YCu$_3$(OH)$_6$Br$_2$[Br$_{1-x}$(OH)$_x$], with their momenta indicated in the kagome Brillouin zone. Adapted from Ref.~\cite{nat.phys.20.1097}. (g) Magnetization of YCu$_3$(OH)$_6$Br$_2$[Br$_{1-x}$(OH)$_x$]. Top: Magnetization curve. Bottom: Differential magnetic susceptibility $dM/dH$. Adapted from Ref.~\cite{pnas.2421390122}.}
    \label{qsl-fig-1}
\end{figure}

The kagome lattice, characterized by its unique geometric configuration where each vertex is shared by two triangles [see Fig.~\ref{qsl-fig-1}(a)], leads to significant frustration for antiferromagnetic (AF) spin interactions. In a typical antiferromagnet, spins align antiparallelly to minimize energy, resulting in an ordered ground state. However, in the kagome lattice, spins cannot simultaneously minimize their interactions with all neighboring spins. As depicted in Fig.~\ref{qsl-fig-1}(a) for the nearest-neighbor (NN) AF Heisenberg model, within each elementary triangle, aligning one spin with a neighbor in an anti-parallel manner creates a situation where the third spin cannot align antiparallelly with all its neighbors. This phenomenon in magnetic systems is called geometric frustration. Strong geometric frustration can melt the magnetic order and is a key factor that may enable kagome AF systems to achieve QSLs. Additionally, the kagome lattice has a low coordination number of 4 for the corner-sharing geometry, which is conducive to realizing QSLs. In contrast,  its triangular counterpart, which also experiences strong geometric frustration, has a coordination number of 6 for the edge-sharing geometry. Extensive theoretical studies have demonstrated that the ground state of the $s=1/2$ NN AF Heisenberg model on a triangular lattice exhibits a 120$^{\circ}$ N\'eel order rather than a QSL~\cite{PhysRevB.50.10048,PhysRevLett.111.157203,PhysRevB.93.144411,PhysRevB.100.241111,PhysRevLett.127.127205,PhysRevLett.127.087201,PhysRevB.108.245102,PhysRevLett.134.196702}, while theoretical studies have consistently shown that the ground state of the $s=1/2$ NN AF Heisenberg model on the kagome lattice is a QSL~\cite{PhysRevB.63.014413,PhysRevLett.98.117205,science.332.1173,PhysRevLett.109.067201,nat.phys.8.902,PhysRevB.87.060405,PhysRevB.91.020402,PhysRevB.95.235107,PhysRevLett.118.137202,PhysRevX.7.031020,Sci.Adv.4.eaat5535,SciPostPhys.7.1.006,PhysRevB.108.L201116,PhysRevB.110.035131,PhysRevLett.133.096501}.

Let's begin by examining the classical ground state of the NN AF Heisenberg model on the kagome lattice. This analysis will reveal why kagome systems are ideal platforms for realizing QSLs. The Hamiltonian is expressed as:
\begin{align}
H=J\sum_{\langle i,j\rangle}\bm{S}_{i}\cdot \bm{S}_{j},
\end{align}
where $J>0$, $\bm{S}_{i}$ represents the spin at site $i$ and $\langle \cdot,\cdot\rangle$ denotes the NN bonds. This Hamiltonian can be rewritten as:
\begin{align}
H=\frac{J}{2}\sum_{\vartriangle,\triangledown}\bm{S}^{2}_{\vartriangle/\triangledown}+\mathrm{const},
\end{align}
where the sum is over all elementary triangles, with $\bm{S}_{\vartriangle/\triangledown}$ being the vector sum of the three spins in a triangle. The energy is minimized when the total spin of each elementary triangle is zero, achieved by arranging all spins at 120$^{\circ}$ angles with their neighbors. Although each triangle's spins lie in a distinct plane, these planes need not align across different triangles due to the kagome lattice's corner-sharing structure, resulting in significant classical ground-state degeneracy~\cite{PhysRevLett.68.855,PhysRevB.47.15342,PhysRevResearch.4.043019}. Let's consider first the subset of ground states where all spins are coplanar. Each coplanar state is highly constrained, featuring only three distinct spin orientations, yet the subset as a whole maintains significant degeneracy. Two typical coplanar states, the $\bm{Q}=0$ state and $\sqrt{3}\times\sqrt{3}$ state are illustrated in Fig.~\ref{qsl-fig-1}(b). Different degenerate coplanar states can be obtained by exchanging two alternating spin orientations along a closed chain containing these two spin orientations. For instance, for the $\bm{Q}=0$ state and the $\sqrt{3}\times\sqrt{3}$ state shown in Fig.~\ref{qsl-fig-1}(b), these chains highlighted by orange shadows are the straight lines along the NN bonds and the elementary hexagons, respectively. Non-coplanar states are formed by rotating the spins of a closed path containing two alternating spin orientations around the axis defined by the spins neighboring the chain. Thus, it is evident that the classical ground state of the NN AF Heisenberg model exhibits extensive degeneracy. This degeneracy is crucial for the potential realization of QSLs, as it prevents the system from settling into a single ordered state. Moreover, the high degeneracy of the classical ground state provides a necessary landscape for quantum fluctuations to act, enabling the system to transition between different degenerate classical ground states and form a QSL. On the other hand, compared to higher spins, the reduced Hilbert space of the $s=1/2$ spin further amplifies quantum fluctuations, making $s=1/2$ kagome antiferromagnets promising platforms for realizing QSLs.

\begin{table}
    \centering
    \caption{Various numerous numerical simulations in literature, including VMC, DMRG, and tensor network methods~\cite{PhysRevLett.98.117205,science.332.1173,PhysRevLett.109.067201,nat.phys.8.902,PhysRevB.87.060405,PhysRevB.91.020402,
PhysRevB.95.235107,PhysRevX.7.031020,SciPostPhys.7.1.006,PhysRevLett.118.137202,Sci.Adv.4.eaat5535,PhysRevB.108.L201116,PhysRevB.110.035131,PhysRevLett.133.096501}, provide increasing evidence suggesting that this QSL is gapless~\cite{PhysRevLett.98.117205,PhysRevB.87.060405,PhysRevB.91.020402,PhysRevX.7.031020,SciPostPhys.7.1.006,PhysRevLett.118.137202,PhysRevB.108.L201116,PhysRevB.110.035131,
PhysRevLett.133.096501} and may even be a Dirac QSL~\cite{PhysRevLett.98.117205,PhysRevB.87.060405,PhysRevB.91.020402,PhysRevX.7.031020,SciPostPhys.7.1.006,PhysRevB.108.L201116,PhysRevB.110.035131,PhysRevLett.133.096501}.}
    \begin{tabular*}{\textwidth}{@{\extracolsep\fill}l c c c}
        \hline \hline
        Method &  Conclusion  &  Ref.  &  Year \\
        \hline
        VMC  &  Dirac QSL  &  \cite{PhysRevLett.98.117205}  &  2007 \\
        DMRG  &  Gapped $\mathbb{Z}_{2}$ QSL  &  \cite{science.332.1173}  &  2011\\
        DMRG  &  Gapped $\mathbb{Z}_{2}$ QSL  &  \cite{PhysRevLett.109.067201}  &  2012\\
        DMRG  &  Gapped $\mathbb{Z}_{2}$ QSL  &  \cite{nat.phys.8.902}  &  2012\\
        VMC  &  Dirac QSL  &  \cite{PhysRevB.87.060405}  &  2013\\
        VMC  &  Dirac QSL  &  \cite{PhysRevB.91.020402}  &  2015\\
        Tensor network  &  Gapped $\mathbb{Z}_{2}$ QSL  &  \cite{PhysRevB.95.235107}  &  2017\\
        Tensor network  &  Gapless QSL  &  \cite{PhysRevLett.118.137202}  &  2017\\
        DMRG  &  Dirac QSL  &  \cite{PhysRevX.7.031020}  &  2017\\
        DMRG  &  Dirac QSL  &  \cite{Sci.Adv.4.eaat5535}  &  2018\\
        Tensor network  &  Dirac QSL  &  \cite{SciPostPhys.7.1.006}  &  2019\\
        VMC  &  Dirac QSL  &  \cite{PhysRevB.108.L201116}  &  2023\\
        VMC  &  Dirac QSL  &  \cite{PhysRevB.110.035131}  &  2024\\
        VMC  &  Dirac QSL  &  \cite{PhysRevLett.133.096501}  &  2024\\
        \hline \hline
    \end{tabular*}
    \label{tab-1:spin_liquid_KAFM}
\end{table}

Theoretically, the $s=1/2$ NN AF Heisenberg model on the kagome lattice has become a cornerstone for research on QSLs, owing to its simplicity and the unique interplay of strong geometric frustration and quantum fluctuations. This model has been extensively investigated through diverse numerical methodologies, including exact diagonalization (ED)~\cite{PhysRevB.83.212401,PhysRevB.100.155142}, variational Monte Carlo (VMC)~\cite{PhysRevLett.98.117205,PhysRevB.87.060405,PhysRevB.91.020402,PhysRevB.108.L201116,PhysRevB.110.035131,PhysRevLett.133.096501}, coupled cluster techniques~\cite{PhysRevB.84.224428,PhysRevB.86.214403}, series expansions~\cite{PhysRevB.76.180407}, density-matrix renormalization group (DMRG)~\cite{science.332.1173,PhysRevLett.109.067201,nat.phys.8.902,PhysRevX.7.031020}, and tensor network~\cite{PhysRevB.95.235107,PhysRevLett.118.137202,SciPostPhys.7.1.006,PhysRevX.4.011025} algorithms. While early studies proposed several candidate ground states, such as a valence-bond crystal with a 36-site unit cell~\cite{PhysRevB.76.180407}, contemporary consensus recognizes a QSL ground state, as summarized in Tab.~\ref{tab-1:spin_liquid_KAFM}.  Notably, variational studies grounded in resonating valence bond theory have provided substantial evidence for a gapless $U(1)$ Dirac spin liquid~\cite{PhysRevLett.98.117205,PhysRevB.87.060405,PhysRevB.91.020402,PhysRevB.108.L201116,PhysRevB.110.035131,PhysRevLett.133.096501}. This scenario is further supported by infinite DMRG simulations~\cite{PhysRevX.7.031020} and tensor network studies~\cite{PhysRevLett.118.137202,SciPostPhys.7.1.006}. Although a recent ED investigation with a 48 site cluster did not conclusively identify characteristics of a gapless $U(1)$ spin liquid, its findings effectively rule out both magnetic ordering and $\mathbb{Z}_{2}$ spin liquid scenarios~\cite{PhysRevB.100.155142}. In contrast, earlier ED~\cite{PhysRevB.83.212401} and DMRG~\cite{science.332.1173,PhysRevLett.109.067201,nat.phys.8.902} studies, alongside variational tensor network calculations~\cite{PhysRevB.95.235107}, have suggested the existence of a gapped spin liquid phase.
The difficulty in accurately determining its ground state primarily stems from the limitations of current computational methods. Among them,  the VMC method provides explicit wave functions for various spin liquids, and enables further studies of their fundmental properties and also serves as a benchmark for other numerical methods. So, on the theoretical side, this chapter will focus on the studies of QSL using the VMC method.

Experimentally, evidences of a QSL state in kagome materials was first observed in herbertsmithite, ZnCu$_3$(OH)$_6$Cl$_2$~\cite{PhysRevLett.98.077204,PhysRevLett.98.107204,Nature.492.7429,science.350.655,Nat.Phys.16.469,PhysRevB.94.060409}, and later in similar compounds such as $\alpha$-Cu$_3$Zn(OH)$_6$Cl$_2$ (kapellasite)~\cite{PhysRevLett.109.037208,PhysRevB.90.205103} and Cu$_3$Zn(OH)$_6$FBr (barlowite)~\cite{CPL.34.077502,PhysRevB.98.155127,CPL.37.107503,npj.QM.5.74,Nat.Commun.12.3048}. In ZnCu$_3$(OH)$_6$Cl$_2$,  for instance, no long-range magnetic order was observed down to low temperatures~\cite{PhysRevLett.98.077204,PhysRevLett.98.107204}, and inelastic neutron scattering revealed a magnetic excitation continuum~\cite{Nature.492.7429,PhysRevB.94.060409}, indicative of the fractional spin excitations characteristic of QSLs. However, the comparable ionic radii of Cu$^{2+}$ and Zn$^{2+}$ (0.73 vs. 0.74 {\AA}) often result in intersite mixing, introducing disorder effects in the kagome lattice~\cite{PhysRevB.90.205103,nat.mater.6.853,PhysRevLett.100.157205}. So, it is challenging to definitively determine whether the ground states of these materials are true QSLs, as the presence of the intersite mixing inevitably influences the low-energy spectra and low-temperature thermodynamic properties. Similar issues have also been found for other QSL candidate materials~\cite{nature.554.341,PhysRevLett.120.087201,nat.commun.9.4367,PhysRevResearch.2.013099,nat.commun.12.4949,PhysRevB.104.224433}.


In recent years, in order to avoid the ionic intersite mixing, a trivalent yttrium ion (Y$^{3+}$) with a larger radius than the Cu$^{2+}$ ion has been introduced to create new systems: YCu$_3$(OH)$_{6+x}$Cl$_{3-x}$~\cite{J.Mater.Chem.C.4.8772,J.Mater.Chem.C.5.2629} and YCu$_3$(OH)$_{6+x}$Br$_{3-x}$~\cite{j.jmmm.2020.167066,PhysRevB.105.L121109,PhysRevB.105.024418}.  In these materials,  magnetic Cu$^{2+}$ ions ($s=1/2$) are expected to fully occupy the regular kagome sites.
Figure~\ref{qsl-fig-1}(c) illustrates the crystal structure of YCu$_3$(OH)$_{6+x}$Br$_{3-x}$, which features a kapellasite-type layer structure with an almost perfect kagome lattice composed of Cu$^{2+}$ ions~\cite{j.jmmm.2020.167066,PhysRevB.105.L121109,PhysRevB.105.024418,pnas.2421390122}. Despite having a Curie–Weiss temperature of about $-80$ K, YCu$_3$(OH)$_{6+x}$Br$_{3-x}$ exhibits no magnetic transition down to 50 mK~\cite{j.jmmm.2020.167066,PhysRevB.105.L121109,PhysRevB.105.024418,PhysRevB.106.L220406,Commun.Phys.5.272}. Moreover, the low-temperature specific heat shows a $T^{2}$  and a $T$-linear dependence at zero and finite field, respectively, which is consistent with a Dirac QSL~\cite{PhysRevB.105.L121109}. Especially, a recent inelastic neutron scattering experiment provides spectral evidence for a Dirac QSL state in this kagome antiferromagnet~\cite{nat.phys.20.1097}. Figure~\ref{qsl-fig-1}(d) shows an energy-momentum plot with momentum $Q$ along the $[H, 0]$ direction. It is evident that there are two filled cone-like excitations at $(2/3, 0)$ and $(4/3, 0)$. The elementary excitations of a QSL are spinons with spin quantum number $s=1/2$, while the $s=1$ magnetic excitations measured in the inelastic neutron scattering experiments are composed of two spinons. Consequently, the spin excitation spectrum theoretically constitutes a continuum of dispersive cones with finite spectral weight spanning from zero energy apex to higher energies [Fig.~\ref{qsl-fig-1}(e) and (f)], fundamentally distinguishing from conventional magnon-like spin waves that exhibit a dispersions with vanishing spectral weight inside the cone. Thus, these experimental observations are consistent with the theoretical expectations for the spectrum of a Dirac QSL. It should be noted that the Dirac QSL state realized in YCu$_3$(OH)$_{6+x}$Br$_{3-x}$ experimentally is not that predicted by the $s=1/2$ NN AF Heisenberg model, which  exhibits conical spin excitations at $(0.5, 0)$~\cite{Proc.Natl.Acad.Sci.116.5437}. This suggests that other exchange interactions beyond the NN Heisenberg interactions in this material may have significant effects.

Very recently, in addition to the evidences for a Dirac QSL, experiments have observed the $1/9$ and $1/3$ magnetization plateaus in YCu$_3$(OH)$_{6+x}$Br$_{3-x}$ and its structurally similar counterpart YCu$_3$(OD)$_{6+x}$Br$_{3-x}$~\cite{nat.phys.20.435,PhysRevLett.132.226701,PhysRevX.15.021076,pnas.2421390122}. Figure~\ref{qsl-fig-1}(g) illustrates the relationship between magnetization ($M$) and magnetic field ($\mu_{0}H$) in YCu$_3$(OH)$_{6+x}$Br$_{3-x}$. For $\mu_{0}H>55$ T, a pronounced magnetization plateau appears around $M=0.35\mu_{B}$ per Cu$^{2+}$ (where $\mu_{B}$ is the Bohr magneton), representing $1/3$ of the Cu$^{2+}$ saturation moment. This indicates a $1/3$ magnetization plateau. The most notable feature is a distinct plateau at $M=0.11\mu_{B}$, indicating a $1/9$ magnetization plateau. The emergence of these plateaus is further supported by the differential magnetic susceptibility $\chi=dM/dH$, which shows two distinct dips at the centers of the $1/9$ and $1/3$ plateaus. Theoretically, both the $1/9$ and $1/3$ plateaus have been discussed as quantum states rather than classical ones found in other frustrated quantum magnets~\cite{nat.com.4.2287,PhysRevLett.133.096501,JPSJ.93.123706,arxiv.2503.11834,cpl.42.090704}. We will provide a detailed introduction to the theoretical research in this area later.

\subsection{Gauge theory, projective construction theory, and variational Monte Carlo}\label{qsl-vmc}

In this subsection, we offer a concise introduction to the VMC method for studying QSLs, along with the associated gauge theory and projective construction theory. Additionally, we discuss how the VMC method can be used to calculate two key quantities: the ground state degeneracy (GSD) and topological entanglement entropy (TEE). These quantities are essential for characterizing the topological excitations and long-range quantum entanglements of gapped QSL states.

In the framework of the projective symmetry group, a crucial aspect is the $SU(2)$ gauge structure of the Abrikosov fermion representation of the $s=1/2$ spin operators, given by $S_{i}^{\alpha} = \psi_{i}^{\dagger} \sigma_{\alpha} \psi_{i}$.  Here, $\psi_{i} = (c_{i,\uparrow}, c_{i, \downarrow})^{T}$, and $\sigma_{\alpha}$ is Pauli matrix with $\alpha \in \{x,y,z\}$. To emphasize the $SU(2)$ gauge structure, the particle-hole partner $\bar{\psi}_{i} = (c_{i,\downarrow}^{\dagger}, -c_{i,\uparrow}^{\dagger})^{T}$ is also used. These can be combined into a matrix $\Psi_{i} = (\psi_{i}, \bar{\psi}_{i})$, allowing the spin operator to be expressed as $S_{i}^{\alpha} = \frac{1}{4} \mathrm{Tr}(\Psi_{i}^{\dagger} \sigma_{\alpha} \Psi_{i})$ or $S_{i}^{\alpha} = \frac{1}{4} \mathrm{Tr}(\Psi_{i}^{\dagger} \Psi_{i} \sigma_{\alpha}^{T})$~\cite{PhysRevB.38.745}. The two expressions are invariant under any local right or left transformation  $\Psi_{i} \rightarrow \Psi_{i} g_{i}$ or $\Psi_{i} \rightarrow g_{i}\Psi_{i}$ for $g_{i} \in SU(2)$, respectively. Moreover, since this fermionic representation expands the physical Hilbert space, a local particle number constraint must be imposed: $N_{i} = \psi_{i}^{\dagger}\psi_{i} = 1$.

For a general spin model, we can rewrite it with doublet operator $\psi_{i}$ and decouple it to the non-interacting mean-field Hamiltonian $H_{\mathrm{mf}}$. This decoupling reduces the $SU(2)$ gauge group to its invariant subgroup, either $U(1)$ or $\mathbb{Z}_{2}$. By considering the symmetry of the lattice geometry and time-reversal transformation, the QSLs can be classified~\cite{PhysRevB.65.165113, PhysRevB.83.224413, PhysRevB.93.094437}. The mean-field Hamiltonian can be rewritten as $H_{\mathrm{mf}} = \sum_{i,j} \mathrm{Tr}(\Psi_{i}^{\dagger} U_{ij} \Psi_{j})$, where $U_{ij} = U_{ji}^{\dagger}$. Introducing the local $SU(2)$ gauge transformation, $\Psi_{j} \to g_{j}\Psi_{j}$, results in $U_{ij} \to g_{i} U_{ij} g_{j}^{\dagger}$. The invariant gauge group (IGG) $\mathcal{G}$ is defined as the subgroup of gauge transformations that leave $U_{ij}$ invariant, i.e., $\mathcal{G} = \{g_{i} U_{ij} g_{j}^{\dagger} = U_{ij}, g_{i} \in SU(2) \}$. Clearly, for $g_{i} = \pm \sigma_{0}$ ($\sigma_{0}$ is a two by two identical matrix), it must lead to the identical transformation $U_{ij} \to U_{ij}$, indicating gauge invariance. That means the IGG for a general $\{ U_{ij} \}$ is at least $\mathbb{Z}_{2}$. However, the IGG can be larger, such as $U(1)$ or even $SU(2)$.

For a 2D system with translational symmetry along two non-collinear lattice vectors $\boldsymbol{a}_{1}$ and $\boldsymbol{a}_{2}$, there is an identical operation given by $T_{2}^{-1} T_{1}^{-1} T_{2} T_{1} = \mathbb{I}$, where $T_{1,2}$ are translation along $\boldsymbol{a}_{1}$ and $\boldsymbol{a}_{2}$, respectively.
In the PSG framework, this identity transformation can be expressed as:
\begin{equation}
    T_{2}^{-1} G_{T_{2}}^{-1} T_{1}^{-1} G_{T_{1}}^{-1} G_{T_{2}} T_{2} G_{T_{1}} T_{1} = \mathcal{G},
\end{equation}
where $G_{T_{1,2}} \in SU(2)$. This leads to:
\begin{equation}
    G_{T_{2}}^{\dagger}(\boldsymbol{r} + \boldsymbol{a}_{2}) G_{T_{1}}^{\dagger}(\boldsymbol{r} + \boldsymbol{a}_{1} + \boldsymbol{a}_{2}) G_{T_{2}}(\boldsymbol{r} + \boldsymbol{a}_{1} + \boldsymbol{a}_{2}) G_{T_{1}}(\boldsymbol{r} + \boldsymbol{a}_{1}) = \mathcal{G}.
\end{equation}
Without loss of generality, we can fix the gauge globally by setting $G_{T_{1}} = \sigma_{0}$. For IGG $\mathcal{G} = \mathbb{Z}_{2}$, there are two global solutions: $G_{T_{2}} = \pm \sigma_{0}$. Especially  the solution $-\sigma_{0}$, which satisfies the magnetic translation algebra, often results in the so-called $\pi$-flux state. This implies that $U_{ij}$ breaks the translational symmetry. However, in the physical Hilbert space with one Abrikosov fermion per site, the corresponding spin wave function retains this symmetry due to the gauge structure, as seen in the Dirac QSL on the kagome lattice~\cite{PhysRevB.83.224413, PhysRevB.93.094437}. For larger symmetry groups, the gauge transformations associated with symmetry operations can be more complex~\cite{PhysRevB.83.224413, PhysRevB.93.094437}. In a word, according to the PSG classification of QSLs, we can explore different types of QSLs by constructing various gauge-inequivalent $\{ U_{ij} \}$.

The mean-field Hamiltonian $H_{\mathrm{mf}}$ can serve as a tool for constructing microscopic wave functions for variational energy calculations in spin models. By utilizing the various gauge non-equivalent Ans\"atze from $H_{\mathrm{mf}}$, the mean-field ground state $|\mathrm{GS}_{\mathrm{mf}} \rangle$ can be constructed. A Gutzwiller projector is then introduced to eliminate double occupancy, yielding the trial many-body wave function $|\Phi(x) \rangle = P_{G}|\mathrm{GS}_{\mathrm{mf}} \rangle$, where $x$ represents various variational parameters. Following this, the projective construction is performed. The trial many-body wave function $|\Phi(x) \rangle$ is related to preliminary preparation of the VMC method, which is one of the powerful computational techniques to study the QSL states based on the projected wavefunctions~\cite{Ann.Phys.189.53}.
The optimal value $x_{\mathrm{opt}}$ can be determined through VMC calculations by minimizing the trial energy $E(x) = \langle \Phi(x) | H | \Phi(x) \rangle / \langle \Phi(x) | \Phi(x) \rangle$. The most probable ground state, with the lowest variational energy, is identified by comparing the optimal energies of various trial states. In principle, all physical quantities of the ground state can be captured by the wave function $|\Phi(x_{\mathrm{opt}}) \rangle = P_{G}|\mathrm{GS}_{\mathrm{mf}} \rangle$. Although the starting point of the VMC technique is based on the solution of the mean-field Hamiltonian, it extends far beyond the mean-field level. Moreover, it is adaptable to various lattice geometries and dimensions, facilitating the exploration of different QSL phases.

In the category of topological order~\cite{Int.J.Mod.Phys.B.4.239, Natl.Sci.Rev.3.68}, the GSD, which refers to the number of degenerate (or quasi-degenerate, in the case of finite systems) ground states, is a key global quantity for characterizing the underlying topological properties of a system. Here, we use the $\mathbb{Z}_{2}$ QSL as an example to illustrate the process for calculating the GSD of a quantum state using the VMC method. For a $\mathbb{Z}_{2}$ QSL state with spinon pairing that cannot be eliminated by any $SU(2)$ gauge transformation, the $\mathbb{Z}_{2}$ gauge-flux excitations are typically gapped. When this QSL state is placed on a torus, inserting a global $\mathbb{Z}_{2}$ flux into any hole of the torus incurs no energy cost, as seen in the canonical toric code model in two dimensions~\cite{ann.phys.303.2}. This operation can be achieved by changing the boundary condition of the mean-field Hamiltonian from periodic to anti-periodic. Typically, one can construct four mean-field ground states with different boundary conditions, denoted as $|\mathrm{GS}_{\mathrm{mf}, \alpha} \rangle$, where $\alpha \in \{++, +-, -+, -- \}$. Here, the sign ``$+$" (``$-$") indicates a periodic (anti-periodic) boundary condition, respectively. However, these four states will vanish after the Gutzwiller projection if the number of unpaired spinons is odd in the construction of $|\mathrm{GS}_{\mathrm{mf}, \alpha} \rangle$. The non-vanishing wave functions after the Gutzwiller projection ($P_{G}$) are denoted as $|\alpha \rangle$. By diagonalizing the overlap matrix of $|\alpha \rangle$, the eigenvalues can be obtained. The number of significantly finite eigenvalues corresponds to the GSD.


Another important topological quantity, TEE, characterizing long-range quantum entanglement in many-body systems, is related to the quantum dimension $d_{i}$ of the topological excitations, and is expressed as $\gamma = \ln D_{q}$, where $D_{q} = \sqrt{\sum_{i} d_{i}^2}$ represents the total quantum dimension of the state~\cite{PhysRevLett.96.110404, PhysRevLett.96.110405}. For Abelian topological orders, where all $d_{i} = 1$, the TEE simplifies to $\gamma = \ln \sqrt{\mathrm{GSD}}$. In contrast, the quantum dimension of non-Abelian quasiparticles exceeds $1$. For example, the Ising anyon, with $d = \sqrt{2}$, appears in the gapless phase of the Kitaev model under a magnetic field~\cite{ann.phys.321.2}.
However, an important robust characteristic, related to the thermal Hall effect, is the chiral central charge $c$ for the gapless edge excitations and not directly captured by the above unitary modular category. Instead, it is determined by the relation:
\begin{equation}
D_{q}^{-1} \sum_{i} d_{i}^2 e^{\mathrm{i} 2\pi s_{i}} = e^{\mathrm{i} 2\pi c / 8},
\end{equation}
where $s_{i}$ is the topological spin~\cite{ann.phys.321.2}, or equivalently, the gravitational Chern-Simons term~\cite{Natl.Sci.Rev.3.68}.

In a general gapped phase of a two-dimensional system, the TEE can be expressed as $S(\mathcal{L}) = \alpha \mathcal{L} - \gamma$ when the system is partitioned into region $A$ and its complement $B$. The area term ($\alpha \mathcal{L}$) dominates in this case, where $\alpha$ is a non-universal factor and $\mathcal{L}$ is the length of the boundary of region $A$, which accounts for the local entanglement at the boundary and determines the scaling of the entanglement. The universal quantity is the TEE $\gamma$.
Since $S(\mathcal{L}) \ge 0$, for a finite $\gamma > 0$, it is impossible to eliminate the area term by topologically deforming the wave function of the state. Conversely, $\gamma = 0$ implies that the gapped state can be deformed into a non-entangled product state. In numerical calculations of entanglement entropy, the second-order Renyi entropy $S^{(2)} = -\ln [\mathrm{Tr} (\rho_{\mathrm{A}}^{2})]$ is commonly used~\cite{PhysRevB.84.075128}, where $\rho_{\mathrm{A}} = \mathrm{Tr}_{\mathrm{B}}(|\Phi\rangle \langle \Phi |)$ is the reduced density matrix of the system described by the wave function $|\Phi \rangle$. To compute $\mathrm{Tr} (\rho_{\mathrm{A}}^{2})$ using the VMC method, an operator $\mathcal{X}_{\mathrm{A}}$~\cite{PhysRevLett.104.157201} is introduced:
\begin{equation}
    \mathcal{X}_{\mathrm{A}} |\alpha\rangle |\beta\rangle = |\alpha^{\prime}\rangle |\beta^{\prime}\rangle,
\end{equation}
where $|\alpha\rangle (|\beta\rangle)= |a_{1}\rangle|b_{1}\rangle$ represents the configuration in lattice space, $c_{r_{1} \sigma_{1}}^{\dagger}...c_{r_{N} \sigma_{N}}^{\dagger}|0\rangle$, with subconfigurations $|a_{1}\rangle$ and $|b_{1}\rangle$ belonging to subsystems $A$ and $B$ respectively.
The operator $\mathcal{X}_{\mathrm{A}}$ swaps parts in the subsystem $A$ between the two configurations, resulting in two new configurations $|\alpha^{\prime}\rangle $ and $|\beta^{\prime}\rangle$. For two copies of the system with the product wave function $|\Phi\rangle |\Phi\rangle$, the Renyi entropy can be rewritten using the expectation value of $\mathcal{X}_{\mathrm{A}}$: $S^{(2)} = -\ln \langle \mathcal{X}_{\mathrm{A}} \rangle$. To conduct the regular Monte Carlo procedure, $\langle \mathcal{X}_{\mathrm{A}} \rangle$ is expressed as:
\begin{equation}
\begin{aligned}
    \langle \mathcal{X}_{\mathrm{A}} \rangle &= \frac{\langle \Phi | \langle \Phi | \mathcal{X}_{\mathrm{A}} | \Phi \rangle | \Phi \rangle}{\langle \Phi | \Phi \rangle \langle \Phi | \Phi \rangle} \\
    &= \sum_{\alpha,\beta} \frac{|\langle \alpha | \Phi \rangle|^{2}}{\langle \Phi | \Phi \rangle} \frac{|\langle \beta | \Phi \rangle|^{2}}{\langle \Phi | \Phi \rangle} \frac{\langle \alpha^{\prime} | \Phi \rangle \langle \beta^{\prime} | \Phi \rangle}{\langle \alpha | \Phi \rangle \langle \beta | \Phi \rangle},
\end{aligned}
\end{equation}
where both $|\langle \alpha | \Phi \rangle|^{2}/\langle \Phi | \Phi \rangle = \rho_{\alpha}$ and $|\langle \beta | \Phi \rangle|^{2}/\langle \Phi | \Phi \rangle = \rho_{\beta}$ are weight factors. The product $\rho_{\alpha}$ and $\rho_{\beta}$ can be treated as the joint distribution probability $\rho_{\alpha, \beta}$ in probability theory, with the sampling of $|\alpha \rangle$ and $|\beta \rangle$ configurations being two independent events. Techniques to improve the calculation of this quantity can be found in Refs.~\cite{PhysRevB.84.075128, PhysRevLett.107.067202, PhysRevB.88.125135}.

In addition,  the braiding rules and self-statistics of the topological excitations have also been used to investigate topological orders using the VMC method~\cite{PhysRevB.85.235151, PhysRevB.87.161113, PhysRevB.97.195158}.

\subsection{Variational Monte Carlo studies of quantum spin liquids in kagome antiferromagnets}\label{sqsh-ekafm}

\begin{figure}
    \centering
    \includegraphics[width=\linewidth]{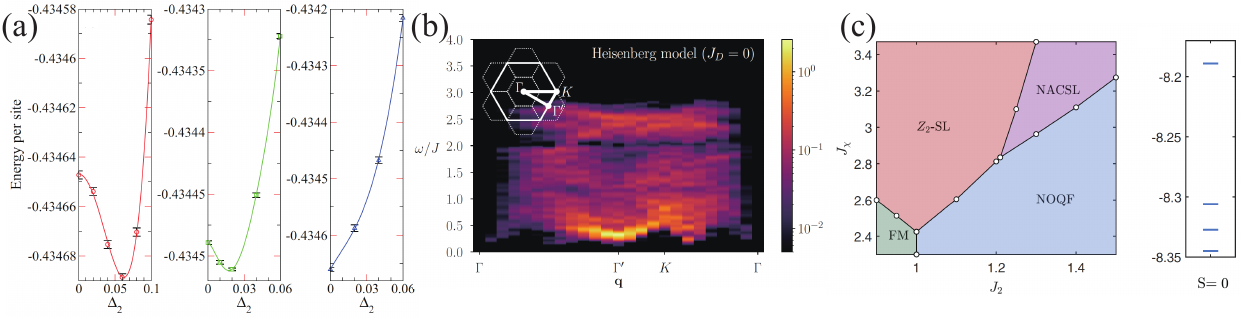}
    \caption{(a) Energy per site of the $\mathbb{Z}_{2}[0, \pi]\beta$ state as a function of the second NN pairing $\Delta_{2}$ when $J_{2}/J_{1} = 0.15$, with system size $L=4, 8$ and $12$ in left, middle and right panels, respectively. Adapted from Ref.~\cite{PhysRevB.91.020402}. (b) Spin dynamical structure factor $S^{z}(\boldsymbol{q}, \omega)$ of the Dirac QSL, calculated by VMC method. Adapted from Ref.~\cite{SciPostPhys.14.139}. (c) Left: Phase diagram of the $s = 1$ $J_{1}$-$J_{2}$-$J_{\chi}$-$J_{r}$ AF Heisenberg model on kagome lattice, including FM order, $Z_{2}$-SL ($Z_{2}$ spin liquid), NOQF (nematic order with quantum fluctuations) and NACSL (non-Abelian chiral QSL). Here, $J_{2}$, $J_{\chi}$ and $J_{r}$ are biquadratic, three-spin chiral and ring exchange interactions, respectively. Right: Three-fold degenerate ground state of NACSL from the ED method. Adapted from Ref.~\cite{PhysRevA.108.033308}.}
    \label{fig:vmc_kafm}
\end{figure}

In this subsection, we focus on the progress made in using VMC methods to study QSLs in kagome antiferromagnets.
The elements of the space group of the kagome lattice include two translations, a $60^\circ$ rotation about the center of a hexagon, and a mirror reflection.
Hence, the PSG analysis suggests two competitive states: a gapless Dirac QSL and a gapped $\mathbb{Z}_{2}$ QSL (referred to as the $\mathbb{Z}_{2}[0,\pi]\beta$ state)~\cite{PhysRevB.83.224413,PhysRevB.96.205150, PhysRevB.93.094437}. The VMC calculations suggest that the Dirac QSL state is favored energetically ~\cite{PhysRevLett.98.117205, PhysRevB.91.020402,PhysRevB.108.L201116,PhysRevB.110.035131,PhysRevLett.133.096501}, and the corresponding mean-field ans\"atz is formed by a $\pi$ flux threading each hexagon breaks translational symmetry, typically doubling the unit cell to six sublattices. The $\mathbb{Z}_{2}[0,\pi]\beta$ state is proposed as the instability resulting from the second-NN spinon pairings, $\Delta_{2}$, in the Dirac QSL state~\cite{PhysRevB.83.224413}. According to the VMC studies~\cite{PhysRevB.91.020402}, the variational energy of this state remains close to that of the Dirac QSL state in small systems, as the second NN AF Heisenberg interaction ($J_{2}$) increases. However, as the system size increases, the spinon pairing $\Delta_{2}$ becomes progressively weaker and eventually vanishes [see Fig.~\ref{fig:vmc_kafm}(a)]. These VMC calculations provide numerical evidence that the Dirac QSL state is the ground state of the kagome NN AF Heisenberg model. Moreover, as shown in Fig.~\ref{fig:vmc_kafm}(b), the dynamical spin structure factor from the VMC calculations exhibits a broad continuum~\cite{SciPostPhys.14.139}, providing a crucial signature for future experimental verifications.
It is noted that recent studies using DMRG~\cite{PhysRevX.7.031020} and tensor network~\cite{PhysRevLett.118.137202} simulations also support a gapless QSL state, with signatures of Dirac cones corresponding to the Dirac QSL~\cite{PhysRevX.7.031020}. While these numerical investigations have provided valuable insights, the exact nature of this QSL is till under investigation.

Beyond the NN Heisenberg exchange interaction, real materials often exhibit other interactions. The study of extended Heisenberg models on kagome lattices has also received widespread attention, leading to the discovery of even more intriguing QSL states. The incorporation of second-NN Heisenberg interactions ($J_{2}$) reveals a rich phase diagram: while the Dirac QSL ground state remains stable for weak $J_{2}$, a phase transition to a $\bm{Q}=0$ magnetically ordered state emerges at sufficiently strong $J_{2}$~\cite{PhysRevB.84.020404,PhysRevLett.108.207204,PhysRevB.91.020402,PhysRevB.89.020408,PhysRevB.91.075112,PhysRevB.91.104418,PhysRevB.95.054418,
PhysRevB.104.144406,PhysRevB.108.184405}. A longer-range interaction, known as the $J_{d}$ term, which links the diagonal spins of the hexagons in the kagome lattice, can stabilize a chiral QSL with spontaneous time-reversal symmetry breaking, as demonstrated by DMRG~\cite{PhysRevB.91.075112} and VMC~\cite{PhysRevB.91.041124} studies. The mean-field ans\"atz of this chiral QSL includes a $\theta_{1}$ in each elementary triangle and a $\pi - 2 \theta_{1}$ flux in each hexagon. A finite gauge flux $\theta_{1}$ can gap the Dirac cones in the Dirac QSL state ($\theta_{1} =0$). A large $J_{d}$ favors a so-called $cuboc1$ magnetic order~\cite{PhysRevLett.108.207204,PhysRevB.91.075112}. It is also proposed that the ground state of the XXZ model with anisotropic $J_{2}$ and $J_{d}$ terms~\cite{PhysRevLett.112.137202, PhysRevLett.114.037201} could be a chiral QSL with semion excitations (topological spin $s = 1/4$) and chiral central charge $c = 1$ resulting in the thermal Hall conductivity $\kappa_{H} = c \frac{\pi}{6} \frac{k_B^2}{\hbar}T = \frac{\pi}{6} \frac{k_B^2}{\hbar}T$~\cite{Natl.Sci.Rev.3.68}.
Thus, it is identified as the lattice version of the $\nu = 1/2$ Kalmeyer-Laughlin state~\cite{PhysRevB.91.041124,PhysRevLett.112.137202,PhysRevLett.114.037201}.
Given that the scalar chiral order parameter $\chi = \langle \boldsymbol{S}_{1} \cdot (\boldsymbol{S}_{2} \times \boldsymbol{S}_{3}) \rangle$ is significant in the chiral QSL, the three-spin chiral interaction $J_{c}=\boldsymbol{S}_{1} \cdot (\boldsymbol{S}_{2} \times \boldsymbol{S}_{3})$ is evidently crucial for stabilizing this chiral QSL~\cite{PhysRevB.91.041124,PhysRevB.92.125122}.

The effects of the Dzyaloshinskii–Moriya (DM) interaction, $H_{D} = D \sum_{\langle ij \rangle} \boldsymbol{D}_{ij} \cdot \boldsymbol{S}_{i} \times \boldsymbol{S}_{j}$, on a kagome lattice have also been explored theoretically~\cite{PhysRevB.95.054418, PhysRevB.66.014422, PhysRevB.87.064423, PhysRevB.92.094433, PhysRevLett.118.267201, PhysRevB.95.014422, PhysRevB.108.144406, SciPostPhys.14.139} and experimentally~\cite{PhysRevLett.101.026405, PhysRevB.83.214406}. The DM interaction favors the $\boldsymbol{Q} = 0$ magnetic order, as shown in Fig.~\ref{qsl-fig-1}(b).
In fact, a strong AF $J_{2}$ term also favors the $\boldsymbol{Q} = 0$ magnetic order~\cite{PhysRevB.91.020402, PhysRevB.91.075112, PhysRevB.95.054418}.
In the $\boldsymbol{Q} = 0$ order, the spin arrangement on the diagonal sites of the hexagon exhibits a ferromagnetic pattern. Therefore, it is expected that an additional AF $J_{d}$ coupling between the diagonal spins can suppress the magnetic order.  It has been shown by the VMC method~\cite{PhysRevB.110.035131} that the $J_{d}$ term suppresses the $\boldsymbol{Q} = 0$ order and leads to the emergence of a QSL after a topological phase transition.  In this QSL, the spin-up and spin-down spinons couple to opposite gauge fields without interacting with each other,  and have opposite Chern numbers ($\pm 1$) for the occupied bands. Thus, edge states carrying opposite spinons move in opposite directions. It shares the same consequence to the quantum spin Hall state~\cite{RevModPhys.83.1057} and is referred as the spinon quantum spin Hall state~\cite{PhysRevB.110.035131}.  The negligible scalar chirality correlation $|\chi_{i}\chi_{j}|$ indicates that the many-body state maintains time-reversal symmetry. Intrinsically, its GSD of $n_{g} = 4$ and TEE of $\gamma \sim \ln 2$ suggest that this state has an Abelian $\mathbb{Z}_{2}$ topological order~\cite{PhysRevB.82.155138}. The detailed properties of the topological excitation warrant further studies using unitary modular theory~\cite{Natl.Sci.Rev.3.68}.

All of the above discussions are based on the analysis of the $s=1/2$ kagome systems. For the higher $s=1$ kagome model, a non-Abelian chiral QSL state with Ising-type anyonic excitations (with a quantum dimension of $\sqrt{2}$) is identified through VMC calculations~\cite{PhysRevA.108.033308}, as shown in the phase diagram in Fig.~\ref{fig:vmc_kafm}(c).  The ED simulation reveals a ground-state degeneracy of 3, which is consistent with the result obtained from the VMC method~\cite{PhysRevA.108.033308, PhysRevB.97.195158}. The realization of non-Abelian anyons holds potential for applications in fault-tolerant topological quantum computation.

\subsection{Magnetic field effects in kagome antiferromagnets}\label{qsl_field}

\begin{figure}
    \centering
    \includegraphics[width=\linewidth]{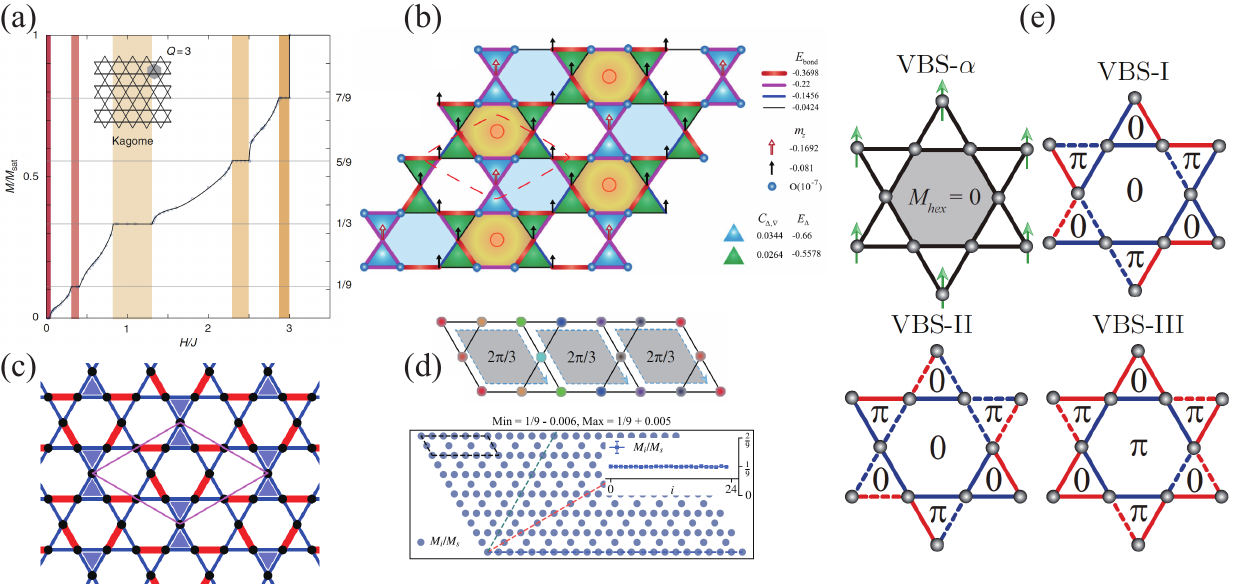}
    \caption{(a) Magnetization curve of the spin-1/2 kagome Heisenberg antiferromagnet in a uniform magnetic field, calculated using the DMRG approach. Five notable magnetization plateaus are observed at values of $M/M_{s}$ ($M_{s} = 1/2$ is the saturated value for the spin-1/2 system): 0, 1/9, 1/3, 5/9, and 7/9,  in addition to the full polarization at $M/M_{s}=1$. Adapted from Ref.~\cite{nat.com.4.2287}. (b) A VBS state proposed by tensor network numerical calculations at the 1/9 plateau. Adapted from Ref.~\cite{PhysRevB.107.L220401}. (c) A $\sqrt{3} \times \sqrt{3}$ VBS state proposed by ED simulations in a small cluster with 36 sites at the 1/9 plateau. Adapted from Ref.~\cite{JPSJ.93.123706}. (d) The ans\"atz with $2\pi/3$ flux threading each primitive cell of kagome lattice and uniform magnetization $M/M_{s} = 1/9$ of $\mathbb{Z}_{3}$ QSL as the ground state of 1/9 magnetization plateau, proposed by VMC study. Adapted from Ref.~\cite{PhysRevLett.133.096501}. (e) The magnetization distribution, labeled by VBS-$\alpha$, from DMRG calculations~\cite{nat.com.4.2287} and the mean-field ans\"atze of the three degenerate $\sqrt{3} \times \sqrt{3}$ VBS states proposed by VMC calculations~\cite{cpl.42.090704}. Red and blue bonds represent two types of hopping terms with different magnitudes, and the dashed bonds represent hopping terms with a minus sign. Adapted from the Ref.~\cite{cpl.42.090704}.}
    \label{fig:magnet_plateau_theory}
\end{figure}

Recent experiments have reported the observation of $1/9$ and $1/3$ magnetization plateaus in YCu$_3$(OH)$_{6+x}$Br$_{3-x}$ and YCu$_3$(OD)$_{6+x}$Br$_{3-x}$~\cite{nat.phys.20.435,PhysRevLett.132.226701,PhysRevX.15.021076,pnas.2421390122}, as shown in Fig.~\ref{qsl-fig-1}(c)-(g). Theoretically, numerous studies have demonstrated that the $s=1/2$ kagome antiferromagnet under an external magnetic field can exhibit magnetization plateaus at 1/9, 1/3 and higher values~\cite{PhysRevLett.88.057204,PhysRevB.70.100403,JPSJ.79.053707,PhysRevB.83.100405,nat.com.4.2287,PhysRevB.88.144416,PhysRevB.93.060407,
PhysRevB.98.014415,PhysRevB.98.094423,nat.com.10.1229,PhysRevB.107.L220401,PhysRevLett.133.096501,JPSJ.93.123706}, as shown in Fig.~\ref{fig:magnet_plateau_theory}(a). This highlights its potential as a promising platform for exploring exotic quantum states, including novel QSLs induced by a magnetic field.

Unlike the commonly observed $1/3$ magnetization plateaus in other antiferromagnets with triangular and honeycomb lattices~\cite{PhysRevB.67.104431,PhysRevB.76.060406,PhysRevLett.108.057205,PhysRevLett.109.267206,PhysRevLett.109.257205,PhysRevLett.110.267201,
nat.com.9.2666,JPSJ.65.2374,nat.phys.19.1883}, which are often characterized by classical spin orders, the $1/9$ and $1/3$ magnetization plateau phases in kagome antiferromagnets are likely exotic quantum states without magnetic order. In particular, the $1/9$ plateau phase is proposed as a $\mathbb{Z}_{3}$ QSL by DMRG~\cite{nat.com.4.2287} and VMC~\cite{PhysRevLett.133.096501} methods. Meanwhile, the tensor network~\cite{PhysRevB.93.060407, PhysRevB.107.L220401} and ED~\cite{JPSJ.93.123706} calculations suggest a $\sqrt{3} \times \sqrt{3}$ valence bond solid (VBS) state at this plateau [see Fig.~\ref{fig:magnet_plateau_theory}(b) and (c)], and  another VMC simulation  proposes a $3\times 3$ VBS state~\cite{arxiv.2507.20308}. In the VMC study~\cite{PhysRevLett.133.096501}, the mean-field ans\"atz and distribution of magnetization of this $\mathbb{Z}_{3}$ QSL are shown in Fig.~\ref{fig:magnet_plateau_theory}(d), respectively. It shows that this QSL state is characterized by a $2\pi/3$ gauge flux threads each primitive cell of the kagome lattice, resulting in a tripling of the unit cell. However, the translational symmetry of the many-body QSL state is preserved due to the characteristic $\mathbb{Z}_{3}$ gauge structure, leading to a uniform magnetization distribution $M_{i}/M_{s} = 1/9$. Further topological information about this disordered state, including the GSD $n_{g} = 9$ and TEE $\gamma = \ln 3$ from VMC calculations~\cite{PhysRevLett.133.096501}, suggests that this state is an Abelian ($\gamma = \ln \sqrt{n_{g}}$) $\mathbb{Z}_{3}$ QSL. Additionally, this state clearly breaks time-reversal symmetry, as indicated by the significant scalar chiral order parameter $|\chi|$. However, the Chern numbers of the occupied bands for opposite spinons are opposite ($C_{\uparrow, \downarrow} = \pm 1$), resulting in a total Chern number of zero but a topologically non-trivial spin Chern number. These physical properties align with the $9_{0}^{B}$ state according to the classification theory of topological order~\cite{Natl.Sci.Rev.3.68}. Thus, such a high-rank topological order is realized in a realistic spin model. This filed-induced $\mathbb{Z}_{3}$ QSL state in the NN AF Heisenberg model is gapped. Furthermore, a recent VMC study shows that an additional DM interaction can reduce the gap of the $\mathbb{Z}_{3}$ QSL and induce a gap closing phase transition in the spinon spectrum~\cite{arxiv.2503.11834}. Currently, there is insufficient experimental evidence to validate various theoretical predictions. Thus, the precise nature of the fascinating 1/9 magnetization plateau phase observed in experiments still requires further theoretical and experimental investigations.

For the 1/3 magnetization plateau observed in experiments at higher magnetic fields~\cite{nat.phys.20.435,PhysRevLett.132.226701,PhysRevX.15.021076,pnas.2421390122}, DMRG~\cite{nat.com.4.2287}, tensor network~\cite{nat.com.10.1229, PhysRevB.93.060407} and VMC~\cite{cpl.42.090704} calculations have identified that this plateau phase corresponds to a $\sqrt{3} \times \sqrt{3}$ VBS state, forming a David-star structure. As shown in the left panel of Fig.~\ref{fig:magnet_plateau_theory}(e), the magnetization distribution of the VBS state (referred to as the ``VBS-$\alpha$" state) from DMRG~\cite{nat.com.4.2287} and tensor network~\cite{nat.com.10.1229} simulations consists of six fully polarized vertices of the David star aligned with the magnetic field direction, and six entangled spins forming a closed loop around the
hexagon with a total $Sz = 0$. This VBS state forms a superlattice with a $\sqrt{3} \times \sqrt{3}$ unit cell, which is distinctly different from the results obtained in other tensor network calculations~\cite{PhysRevB.93.060407} and recent VMC studies~\cite{cpl.42.090704}. The VMC simulation identifies three degenerate $\sqrt{3} \times \sqrt{3}$ VBS states (labeled VBS-I, -II and -III) as the ground states of the 1/3 plateau. The magnetization patterns are also of David-star structures, but have only two fractional values of magnetization, $M_{i}/M_{s} = -1/3$ and $2/3$.  It is worth noting that a similar result, with only one positive and one negative value, was proposed by tensor network studies~\cite{PhysRevB.93.060407}. Additionally, although the spin correlations in the three VBS states are short-range, resembling a QSL, they are topologically trivial many-body states, characterized by a vanishing topological entanglement entropy in the VMC calculation~\cite{cpl.42.090704}.

It has been shown that a magnetization plateau for a spin-$S$ can emerge when the condition $nS(1 - M/S) =\mathbb{Z}$ is satisfied~\cite{nat.com.4.2287, PhysRevLett.78.1984, PhysRevB.79.064412}, where $n$ represents the number of sites in a unit cell of the ground state, $M/S$ is the magnetization, and $\mathbb{Z}$ is an integer. For the 1/3-magnetization plateau phase discussed above, $n=9$, so the value of $nS(1 - M/S) = \frac{9}{2}(1 - \frac{1}{3}) $ is 3, which is an integer, aligning with previous classical arguments on magnetization plateaus~\cite{nat.com.4.2287, PhysRevLett.78.1984, PhysRevB.79.064412}. The three VBS states from the VMC calculations, featuring a $\mathbb{Z}_{1}$ gauge structure because of zero TEE, are consistent with these general arguments. For the $\mathbb{Z}_{3}$ QSL state without translational symmetry breaking (i.e., $n = 3$) at the 1/9-magnetization plateau~\cite{nat.com.4.2287, PhysRevLett.133.096501}, $nS(1 - M/S) = \frac{3}{2}(1 - \frac{1}{9}) = 1 + 1/3$, due to the $\mathbb{Z}_{3}$ gauge field~\cite{PhysRevB.79.064412}. Additionally, for other states with a $\mathbb{Z}_{2}$ gauge field, the quantity is modified to $nS(1 - M/S) = \mathbb{Z} + 1/2$ (where $\mathbb{Z}$ is an integer), as proposed by the effective field theory~\cite{PhysRevB.79.064412}. These results may suggest a conjecture $nS(1 - M/S) = \mathbb{Z} + 1/q$, where $q$ comes from  the $\mathbb{Z}_{q}$ gauge structure of  many-body quantum states. This is an intriguing issue that warrants further study.

\section{Kagome superconductivity}\label{sec-sc}

\subsection{Basic properties of electron structures}\label{sec-sc-1}

\begin{figure}
	\centering
	\includegraphics[width=1.0\linewidth]{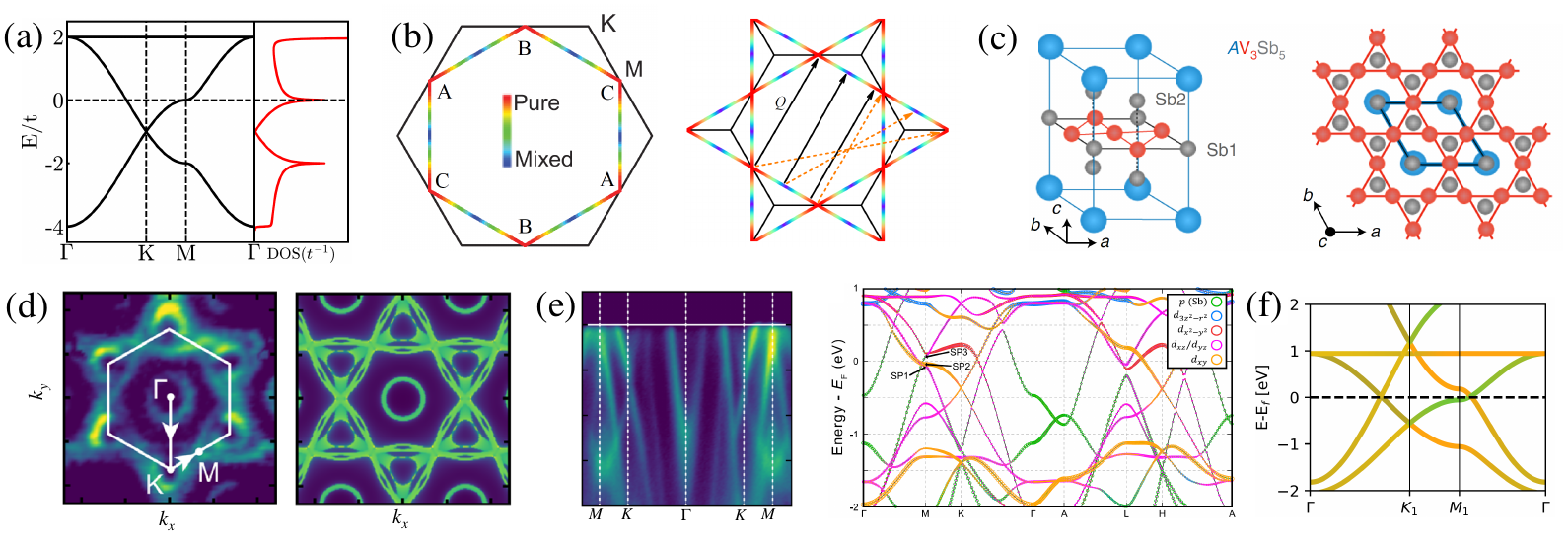}
	\caption{(a) Band structures for the NN tight-binding model on the kagome lattice. Adapted from Ref.~\cite{PhysRevB.85.144402}. (b) Left: Fermi surface at the upper van Hove filling. The colors represent the sublattice weights. Adapted from Ref.~\cite{PhysRevB.85.144402}. Right: Nesting properties of Fermi surface. The solid black lines with arrows represent the nesting property for particle-hole scattering, while the orange dashed lines with arrows indicate the nesting property for particle-particle scattering. For clarity, the nesting wave vector in the particle-particle channel is adjusted from $\bm{Q}$ to $\bm{Q}+\bm{G}$, where $\bm{G}$ is a reciprocal lattice vector. (c) Left: Crystal structure of AV$_3$Sb$_5$ (A=K,Rb,Cs). Right: Projection of the crystal structure along the $c$-direction, showing V atoms forming a perfect kagome lattice. Adapted from Ref.~\cite{Nat.Phys.18.137}. (d) ARPES-measured Fermi surfaces (left) compared with DFT calculations (right) for CsV$_3$Sb$_5$. Adapted from Ref.~\cite{PhysRevLett.125.247002}. (e) Left: ARPES-measured band structures. Adapted from Ref.~\cite{PhysRevLett.125.247002}. Right: Band structures from DFT calculations for CsV$_3$Sb$_5$. Adapted from Ref.~\cite{PhysRevB.105.235145}. (f) Band structures of a minimal two-orbital tight-binding model. Adapted from Ref.~\cite{PhysRevLett.127.177001}.}
	\label{sc-fig-1}
\end{figure}

The kagome lattice is not only an ideal platform for realizing QSLs due to its unique geometric structure, but also possesses an electronic structure with distinctive properties. These properties allow itinerant electronic systems on the lattice to exhibit a variety of exotic quantum states of matter~\cite{PhysRevB.80.113102,PhysRevB.80.193304,PhysRevB.82.075125,JPCM.23.175702,PhysRevLett.106.236802,PhysRevB.85.144402,PhysRevB.87.115135,
	PhysRevB.86.121105,PhysRevLett.110.126405,PhysRevLett.115.186802}. Firgure~\ref{sc-fig-1}(a) illustrates the band structure of a NN tight-binding model on the kagome lattice:
\begin{align}
	H_0=-t\sum_{\langle ij\rangle,\sigma}(c^{\dag}_{i\sigma}c_{j\sigma}+h.c.),
\end{align}
where $c^{\dag}_{i\sigma}$ ($c_{i\sigma}$) creates (annihilates) an electron with spin $\sigma$ at site $i$ and $\langle ij\rangle$ denotes the NN bond. The electronic band structure features an exactly flat band at energy $E=2t$, resulting from the complete destructive interference of electron hopping due to the lattice's unique geometry. Additionally, the band structure includes standard Dirac cones at the $K$ points in the Brillouin zone, similar to those in graphene~\cite{RevModPhys.81.109}. Moreover, there are two van Hove singularities at energies $E=0$ and $E=-2t$ located at the $M$ points, leading to two pronounced peaks in the density of states (DOS)~\cite{PhysRevB.85.144402}. Relative to half filling, the Fermi levels at $E=0$ and $E=-2t$ correspond to the $1/6$ and $1/2$ hole doping, respectively. Interestingly, the Fermi surfaces (FS) at the van Hove fillings are perfectly nested [see Fig.~\ref{sc-fig-1}(b)]. These unique properties have been shown to lead to various instabilities of the normal state at van Hove filling, depending on different interactions, resulting in intriguing quantum phases such as unconventional superconductivity, spin density waves, and charge density waves~\cite{PhysRevB.85.144402,PhysRevB.87.115135,
PhysRevB.86.121105,PhysRevLett.110.126405}. In recent years, the van Hove singularity has been considered a potential driving source for superconductivity and charge density waves in the kagome materials AV$_3$Sb$_5$ (A=K,Rb,Cs)~\cite{PhysRevLett.125.247002,Nat.Phys.18.137,CPB.31.097405,nature.612.647,nsr.10.nwac199,nat.rev.phys.5.635,nat.rev.mat.9.420,nat.phys.20.534}. Consequently, the study of physics related to van Hove filling in the kagome lattice has become a hot topic in condensed matter physics.

For itinerant electronic systems, the minimal description naturally begins with the Hubbard model, rather than the Heisenberg model discussed above in Sec. 2.  This model
is defined as,
\begin{align}
H=H_0+U\sum_{i}n_{i\uparrow}n_{i\downarrow}-\mu \sum_{i,\sigma} n_{i\sigma},
\end{align}
where,  $n=c^{\dag}_{i\sigma}c_{j\sigma}$, $U$ and $\mu$ are the on-site Coulomb interaction and  chemical potential, respectively.   In this weak coupling  framework, various charge orders, spin orders, and SC pairings are governed by the Bethe-Salpeter equations for particle-hole and particle-particle susceptibilities. We will first briefly analyze the characteristics of these susceptibilities, stemming from the unique electronic structure of the kagome lattice at van Hove fillings. In a system with spin rotational invariance, the collective fluctuations—which signal potential instabilities of the normal state—can be classified by the total spin $S$ of the underlying two-particle operators.  The $S=0$ and $S=1$ particle-hole channels correspond to a charge density wave  and a spin density wave, respectively. In the particle-particle channel, the $S=0$ and $S=1$ channels correspond to spin-singlet and spin-triplet superconductivity, respectively.

In the particle-hole channel,  the susceptibility will be greatly enhanced and exhibit a strong peak at a wave vector $\bm{Q}$ if the Fermi surface possesses parallel segments, allowing many pairs of electron states to be connected by the same vector $\bm{Q}$, a phenomenon known as nesting. The FSs at the upper van Hove filling of a kagome lattice, as shown in Fig.~\ref{sc-fig-1}(b),  have perfect nesting properties. In the case of a simple lattice with a single atom in the primitive unit cell, this perfect nesting will lead to the divergence of the particle-hole susceptibility and the formation of density waves with an infinite interaction $U$. However, the kagome lattice is a composite lattice with three atoms (three inequivalent lattice sites) in the unit cell. In Ref.~\cite{PhysRevB.85.144402}, Yu and Li point out that the nesting FS for the upper van Hove filling at $1/6$ hole doping exhibits a unique mixture of the contributions from the three inequivalent lattice sites, and this sublattice characters of the FSs play a vital role in determining the spin, charge and SC orders, which was also emphasized in the subsequent functional renormalization group studies~\cite{PhysRevB.87.115135,PhysRevB.86.121105,PhysRevLett.110.126405}. As depicted in Fig.~\ref{sc-fig-1}(b), the eigenstates at the Fermi level originate exclusively from a single sublattice only at the saddle points ($M$ points). At all other Fermi momenta, the eigenstates form a superposition of two sublattice sites, with their relative weights varying with momentum. As a result,  the matrix element effect considerably weakens the nesting effect in the particle-hole channels for an on-site Hubbard interaction $U$, leading to the appearance of a spin density wave only at a relative large  $U$ and leaving a space for a chiral  $d+id$ SC state realized at a small to intermidiate $U$~\cite{PhysRevB.85.144402}.

Superconductivity arises in the particle-particle channels, where the center-of-mass momentum $\bm{q}$ connects the two single-particle states $|\bm{k}\rangle$ and $|-\bm{k}+\bm{q}\rangle$. This is distinct from the particle-hole case, where  $\bm{q}$ connects the states $|\bm{k}\rangle$ and $|\bm{k}+\bm{q}\rangle$. Commonly, the superconductivity involves the $\bm{q}=0$ pairings,  with a Cooper pair composed of two electrons with momenta $\bm{k}$ and  $-\bm{k}$.  In a kagome lattice at van Hove filling,  the two electrons in a Cooper pair (with momenta  $\bm{k}$ and  $-\bm{k}$) can each reside at one of the two van Hove singularities within the same sublattice (A, B or C in Fig.~\ref{sc-fig-1}(b)). This configuration leverages the exceptionally high density of states provided by the vHS, and is shown to significantly enhance the SC pairing made from two electrons belonging to the same sublattice~\cite{PhysRevB.85.144402}.  As the effective lattice for each sublattice is the triangular lattice,  a $d+id$ wave pairing is favored resulting from the NN pairings with a phase $0, 2\pi/3, 4\pi/3$ along the three NN bonds~\cite{PhysRevB.85.144402}. Later on, it is shown that,  even for on-site pairings, the system may also support a $d+id$ wave in addition to the conventional $s$ wave due to its special three-sublattice structure~\cite{PhysRevB.108.144508}. On the other hand, in kagome lattices at van Hove fillings, there is also a $\bm{q}=\bm{Q}$ nesting wave vector [Fig.~\ref{sc-fig-1}(b)], leading to SC instability with finite momentum and resulting in a pairing-density-wave state~\cite{PhysRevB.111.094505}. Additionally, the unique sublattice distributions on the FSs can further diversify potential SC states. Consequently, van Hove filling kagome systems can realize various exotic SC states, with the specific state achieved depending on the form of the interaction~\cite{PhysRevB.85.144402,PhysRevB.87.115135,PhysRevB.86.121105,PhysRevLett.110.126405,PhysRevB.104.035142,Nat.Commun.13.7288,PhysRevB.108.L081117,PhysRevB.109.075130,PhysRevB.111.094505}.

The previous discussions on SC pairings mainly rely on weak-coupling theories, which emphasize the role of the electronic structure of the kagome system. However, in the context of kagome superconductivity, strong-coupling approaches offer a complementary perspective by highlighting the importance of local electron correlations. Within strong-coupling theories, superconductivity is often obtained by doping spin liquids, typically through doped $t$-$J$ models. Using slave-boson and Chern–Simons formulations, Ko \textit{et al}. proposed a SC state that breaks time-reversal symmetry~\cite{PhysRevB.79.214502} and showed that the flux carried by a minimal vortex is $hc/4e$. In this scenario, the system hosts fermionic quasiparticles with semionic mutual statistics and bosonic $s=1/2$ quasiparticles, in addition to the conventional $s=1/2$ fermionic Bogoliubov quasiparticles. In a VMC study, Jiang \textit{et al}. explored the $SU(2)$ gauge structure hidden in the projective construction at half-filling~\cite{PhysRevLett.127.187003}. By treating gauge-rotation angles as variational parameters, they identified a chiral, noncentrosymmetric nematic SC state, which corresponds to a pair-density-wave state with a $2\times2$ enlarged unit cell and a finite Fermi surface for the Bogoliubov quasiparticles~\cite{PhysRevLett.127.187003}.

The recently discovered SC materials AV$_3$Sb$_5$~\cite{PhysRevLett.125.247002,Nat.Phys.18.137,CPB.31.097405,nature.612.647,nsr.10.nwac199,nat.rev.phys.5.635,nat.rev.mat.9.420} are specific material realizations of the van Hove filling in kagome systems. As illustrated in Fig.~\ref{sc-fig-1}(c), these materials possess a layered structure with a $P6/mmm$ space group, where the V-Sb layers are intercalated with K/Rb/Cs atoms. Within the V-Sb layer, the V atoms form a kagome lattice, intertwined with a Sb triangular lattice and sandwiched between two Sb honeycomb sheets.  Angle-resolved photoemission spectroscopy (ARPES) experimental measurements and density functional theory (DFT) calculations have provide direct information for us to understand the electronic natures of these materials. Figure~\ref{sc-fig-1}(d) displays the FSs of CsV$_3$Sb$_5$ obtained from ARPES experiments and DFT calculations~\cite{PhysRevLett.125.247002}, demonstrating excellent agreement. A notable feature is their nearly hexagonal FSs, which closely resemble the FS of the van Hove filling NN kagome tight-binding model discussed above. This suggests that FS nesting and van Hove singularity are crucial for understanding the nature of charge density waves and superconductivity in these materials. The band structures from ARPES~\cite{PhysRevLett.125.247002} and DFT~\cite{PhysRevB.105.235145} are shown in Fig.~\ref{sc-fig-1}(e), also demonstrating excellent agreement. Besides the hexagonal FSs related to van Hove filling, there is an approximately circular FS around the $\Gamma$ point. According to DFT, the pocket near the $\Gamma$ point originates from the Sb $5p$ orbitals, while the pockets associated with van Hove filling are composed of the V $3d$ orbitals. These materials exhibit charge density waves above the SC transition temperature and feature a $2\times2$ structure in the kagome plane, consistent with the nesting wave vectors shown in Fig.~\ref{sc-fig-1}(b). Thus, a minimal model containing the V-$3d$ orbitals is expected to capture the main physics of these materials. Figure~\ref{sc-fig-1}(f) shows the band structure of a two-orbital tight-binding model, which approximates the electronic structure near the Fermi level in Fig.~\ref{sc-fig-1}(e). From Fig.~\ref{sc-fig-1}(e) and (f),  it is evident that, due to its multi-orbital nature, the two van Hove singularities in the kagome system coexist near the Fermi level of AV$_3$Sb$_5$. This may play a significant role in the symmetry breaking of AV$_3$Sb$_5$, even potentially leading to chiral exciton order~\cite{Nat.Commun.14.605}. Additionally, a theoretical examination of the band structure of CsV$_3$Sb$_5$ uncovers several topologically nontrivial crossings between neighboring bands within the energy range of $\pm1$ eV from the Fermi level, positioning CsV$_3$Sb$_5$ as a $\mathbb{Z}_{2}$ topological metal~\cite{PhysRevLett.125.247002}.

\subsection{Phase diagram, spin and gap structures of the superconducting states in AV$_3$Sb$_5$}

\begin{figure}
    \centering
    \includegraphics[width=1.0\linewidth]{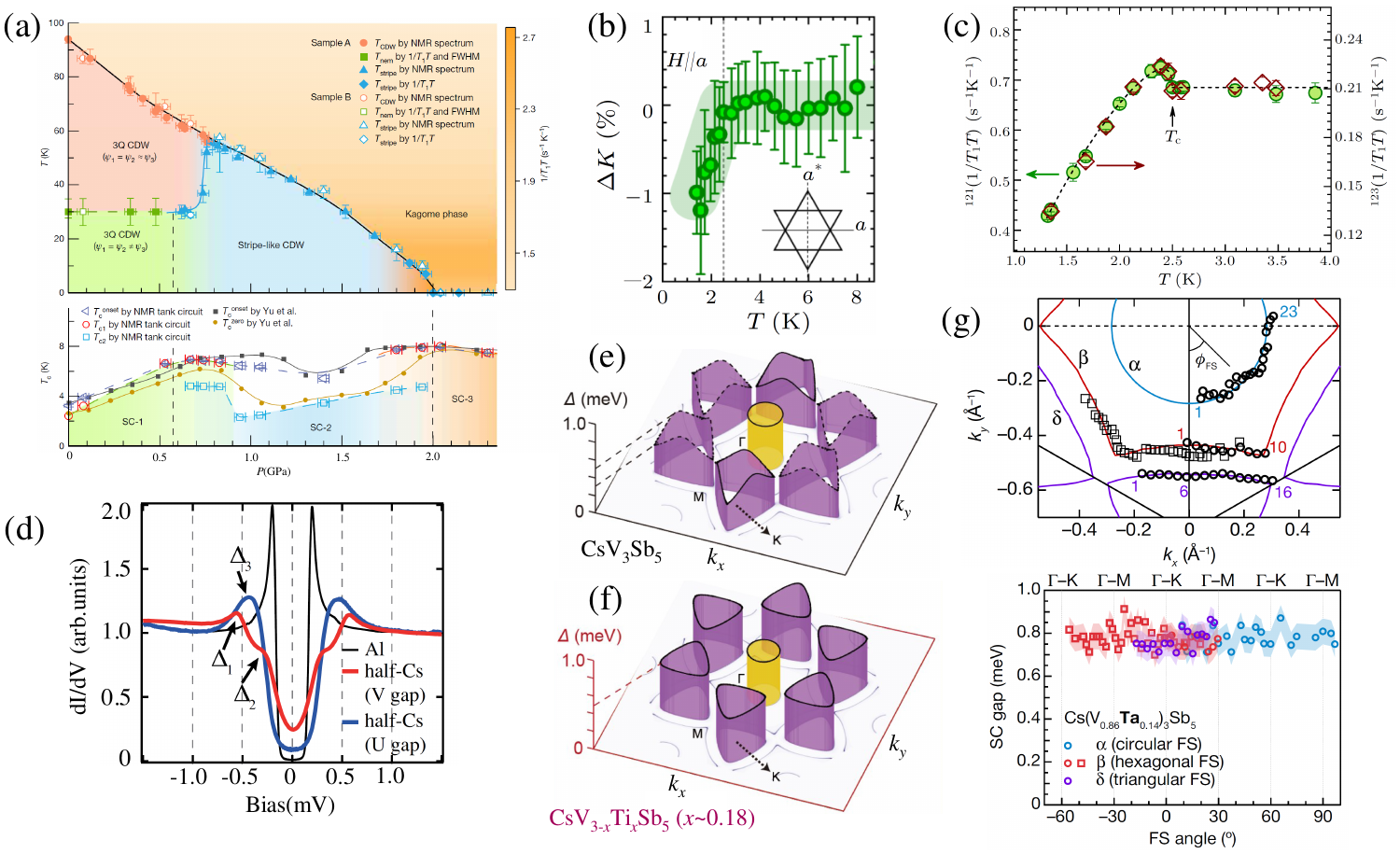}
    \caption{(a) Pressure-temperature phase diagram of CsV$_3$Sb$_5$ determined by $^{51}$V NMR measurements.
    	Adapted from Ref.~\cite{nature.611.682}. (b) Temperature dependence of the Knight shift $\Delta K$ of $^{121}$Sb for CsV$_3$Sb$_5$ with $H\| a$. Adapted from Ref.~\cite{cpl.38.077402}. (c) Temperature dependence of $^{121}$($1/T_1T$) (left axis) and $^{123}$($1/T_1T$) (right axis) for CsV$_3$Sb$_5$. A Hebel–Slichter coherence peak appears just below $T_{\mathrm{c}}$. Adapted from Ref.~\cite{cpl.38.077402}. (d) Two kinds of SC spectra measured on half-Cs surface for CsV$_3$Sb$_5$. SC spectrum of Al polycrystal ($T_{\mathrm{c}}\sim1.2$ K)  is displayed for comparison. Adapted from Ref.~\cite{PhysRevLett.127.187004}. (e), (f) SC energy gap structures measured by Bogoliubov quasiparticle interference for CsV$_3$Sb$_5$ and CsV$_{3-x}$Ti$_x$Sb$_5$ ($x\sim0.18$). Adapted from Ref.~\cite{SCPMA.67.277411}. (g) SC energy gap structures measured by ARPES for Cs(V$_{0.86}$Ta$_{0.14}$)$_3$Sb$_5$. Top: Positions of the Fermi wave vectors. Bottom: Gap amplitudes. Adapted from Ref.~\cite{nature.617.488}.}
    \label{sc-fig-2}
\end{figure}

Figure~\ref{sc-fig-2}(a) presents the pressure-temperature phase diagram for CsV$_3$Sb$_5$, featuring the highest SC transition temperature $T_c$ among the pristine members of  AV$_3$Sb$_5$, based on $^{51}$V nuclear magnetic resonance (NMR)~\cite{nature.611.682} and transport~\cite{Nat.Commun.12.3645} measurements. The phase diagram includes three charge density wave (CDW) phases and three SC phases. At ambient pressure, this material undergoes a $2\times2$ CDW transition at 94 K~\cite{PhysRevLett.125.247002}. This CDW maintains the rotational symmetry of the kagome lattice and develops at all three nesting wave vectors, thus it is also referred to as a triple-$Q$ CDW order~\cite{nature.611.682,PhysRevB.104.045122,PhysRevB.109.104512}. As the temperature decreases, a new electronic nematicity is developed below a characteristic temperature $T_{\mathrm{nem}}\sim35$ K~\cite{nature.611.682,nature.604.59}. Theoretically, this transition from a triple-$Q$ CDW order to a nematic CDW state can be explained by a competitive scenario, where the interplay between doping deviation from the van Hove filling and thermal broadening of the FS determines the energetically preferred state~\cite{PhysRevB.109.104512}. With pressure increasing beyond $P_{c1}\approx0.58$ GPa, a stripe-like CDW order arises from the electronic nematic state through a first-order phase transition. Eventually, this stripe-like CDW is completely suppressed at $P_{c2}\approx2$ GPa through another first-order phase transition, returning the system to an undistorted kagome lattice. In contrast to the decrease in CDW transition temperature with pressure, the SC transition temperature $T_{\mathrm{c}}$ initially increases with pressure, reaching a peak at $P_{1}\approx0.7$ GPa. Beyond this point, superconductivity is suppressed until around $1.1$ GPa. Afterward, $T_{\mathrm{c}}$ rises again as pressure increases. The CDW is fully suppressed at a critical pressure of $P_{2}\approx2$ GPa, where $T_{\mathrm{c}}$ reaches a maximum of approximately $8$ K. This phase diagram reflects a complex interplay between superconductivity and CDW states in these kagome superconductors. Interestingly, growing experimental evidence suggests that the CDW state in AV$_3$Sb$_5$ may exhibit another significant anomalous property: the breaking of time-reversal symmetry~\cite{Nat.mater.20.1353,Nature.602.245,Nat.Phys.18.1470,Nature.611.461,Nat.Phys.20.40,Nature.631.60}. While excellent reviews have extensively covered the unconventional CDW state~\cite{Nat.Phys.18.137,CPB.31.097405,nature.612.647,nsr.10.nwac199,nat.rev.phys.5.635,nat.rev.mat.9.420},  we will focus the following discussion on recent experimental advances in the study of its SC states.

Nuclear magnetic resonance (NMR) experiments provide key evidence for spin-singlet pairing in kagome superconductors. The measured Knight shift in CsV$_3$Sb$_5$ exhibits a clear decrease as the temperature drops below $T_{\mathrm{c}}$~\cite{cpl.38.077402} as shown in Fig.~\ref{sc-fig-2}(b). This drop is a characteristic signature of spin-singlet pairing, as it indicates a reduction in the spin susceptibility of the SC condensate. This behavior stands in direct contrast to that expected for a spin-triplet pairing, in which the Knight shift would typically remain relatively unchanged across $T_{\mathrm{c}}$ due to the presence of parallel spins in the Cooper pairs.
NMR experiments can also provide information about SC gap structures by measuring the spin-lattice relaxation rate $1/T_{1}T$. As shown in Fig.~\ref{sc-fig-2}(c), $1/T_{1}T$ exhibits a clear Hebel-Slichter coherence peak just below $T_{\mathrm{c}}$ and then rapidly decreases at low temperatures~\cite{cpl.38.077402}. Since this coherence peak is typically associated with $s$-wave superconductors, it was proposed that the SC gap may have a $s$-wave symmetry~\cite{cpl.38.077402}. Meanwhile, a scanning tunneling microscopy (STM) measurement on CsV$_3$Sb$_5$ reveals two distinct SC gaps with multiple sets of coherent peaks [see Fig.~\ref{sc-fig-2}(d)]. This observation aligns with the multiple FSs identified by DFT calculations and ARPES measurements, as discussed earlier. However, likely due to varying STM tip conditions across different measurements, the SC gap appears either V-shaped or U-shaped with a small residual DOS, on both the half-Cs and Sb surfaces~\cite{PhysRevLett.127.187004}. Additionally, the in-gap states are induced by magnetic impurities like Cr but not by nonmagnetic impurities like Zn, suggesting a sign-preserving or $s$-wave SC order parameter~\cite{PhysRevLett.127.187004}. The residual spectral weights at zero energy may originate from nodal structures of SC gaps on some Fermi pockets, possibly due to the modulation by CDW order~\cite{PhysRevB.109.104512} or the formation of a pairing density wave~\cite{Nature.632.775}. Furthermore, by implementing Bogoliubov quasiparticle interference imaging, the momentum-space structure of SC gaps can also be determined through STM experiments. As shown in Fig.~\ref{sc-fig-2}(e), in pristine CsV$_3$Sb$_5$, where superconductivity coexists with CDW, the estimated gap on the pocket of V $3d$ orbitals is anisotropic but nodeless, with a minimal value located near the $M $ point, while the gap on the pocket of Sb $5p$ orbitals is nearly isotropic~\cite{SCPMA.67.277411}. For CsV$_{3-x}$Ti$_x$Sb$_5$ ($x\sim0.18$) [Fig.~\ref{sc-fig-2}(f)], where the long-range CDW order is suppressed by replacing V with Ti, nearly isotropic gaps on both the V and Sb pockets are found~\cite{SCPMA.67.277411}. This is consistent with ARPES experiment results in Cs(V$_{0.86}$Ta$_{0.14}$)$_3$Sb$_5$ [see Fig.~\ref{sc-fig-2}(g)]~\cite{nature.617.488}, where the CDW order is also suppressed and the SC gap is nearly isotropic. These observations indicate that without CDW, the SC gap is nearly isotropic, and the CDW strongly affects the gap.

Based on the experimental results introduced above from NMR, STM, and ARPES, it appears that the kagome materials AV$_3$Sb$_5$ exhibit conventional $s$-wave superconductivity. However, recent theoretical studies suggest caution when applying standard interpretations to these experimental findings. The sublattice weight of the band eigenstates disscused in Sec.~\ref{sec-sc-1} plays a crucial role in determining the physical responses related to superconductivity on the kagome lattice.  For instance, although the unconventional gap function averages to zero over the FS ($\sum_{\bm{k}}\Delta(\bm{k})=0$), the atomic-scale disorder limits scattering between Bloch eigenstates to specific regions of the FS. This results in the absence of in-gap bound states in spin-singlet superconductors for nonmagnetic disorders~\cite{PhysRevB.108.144508}. This behavior contrasts with the $d$-wave pairing on a square lattice, where nonmagnetic disorder induces distinct impurity resonant states~\cite{PhysRevB.51.15547}. Moveover, for unconventiaonal $d$-wave pairings on the kagome lattice, regardless of whether time-reversal symmetry is broken, the spin-lattice relaxation rate shows a pronounced Hebel-Slichter peak upon entering the SC state~\cite{PhysRevB.106.014501,PhysRevB.110.144516}. Thus, although many experimental results point to characteristics reminiscent of conventional $s$-wave superconductivity, a conclusive identification of AV$_3$Sb$_5$ as conventional superconductors remains elusive. In fact, several experiments have revealed that the SC state of kagome materials AV$_3$Sb$_5$ exhibits a variety of unconventional properties. For instance, the SC state intertwines with several exotic electronic orders, including time-reversal-symmetry-breaking charge orders, nematicity, and topological states~\cite{PhysRevLett.125.247002,Nat.Phys.18.137,CPB.31.097405,nature.612.647,nsr.10.nwac199,nat.rev.phys.5.635,nat.rev.mat.9.420}. Notably, magnetoresistance oscillation measurements suggest the possible emergence of exotic $4e$ and $6e$ pairings in the superconducting ring of CsV$_3$Sb$_5$~\cite{PhysRevX.14.021025}. In the following sections, we focus on two key aspects of SC state: time-reversal symmetry breaking and translational symmetry breaking, which are important for understanding the nature of superconductivity in AV$_3$Sb$_5$ and for uncovering its underlying mechanism.

\subsection{Experimental evidences of chiral superconductivity and pairing density wave}

\begin{figure}
    \centering
    \includegraphics[width=1.0\linewidth]{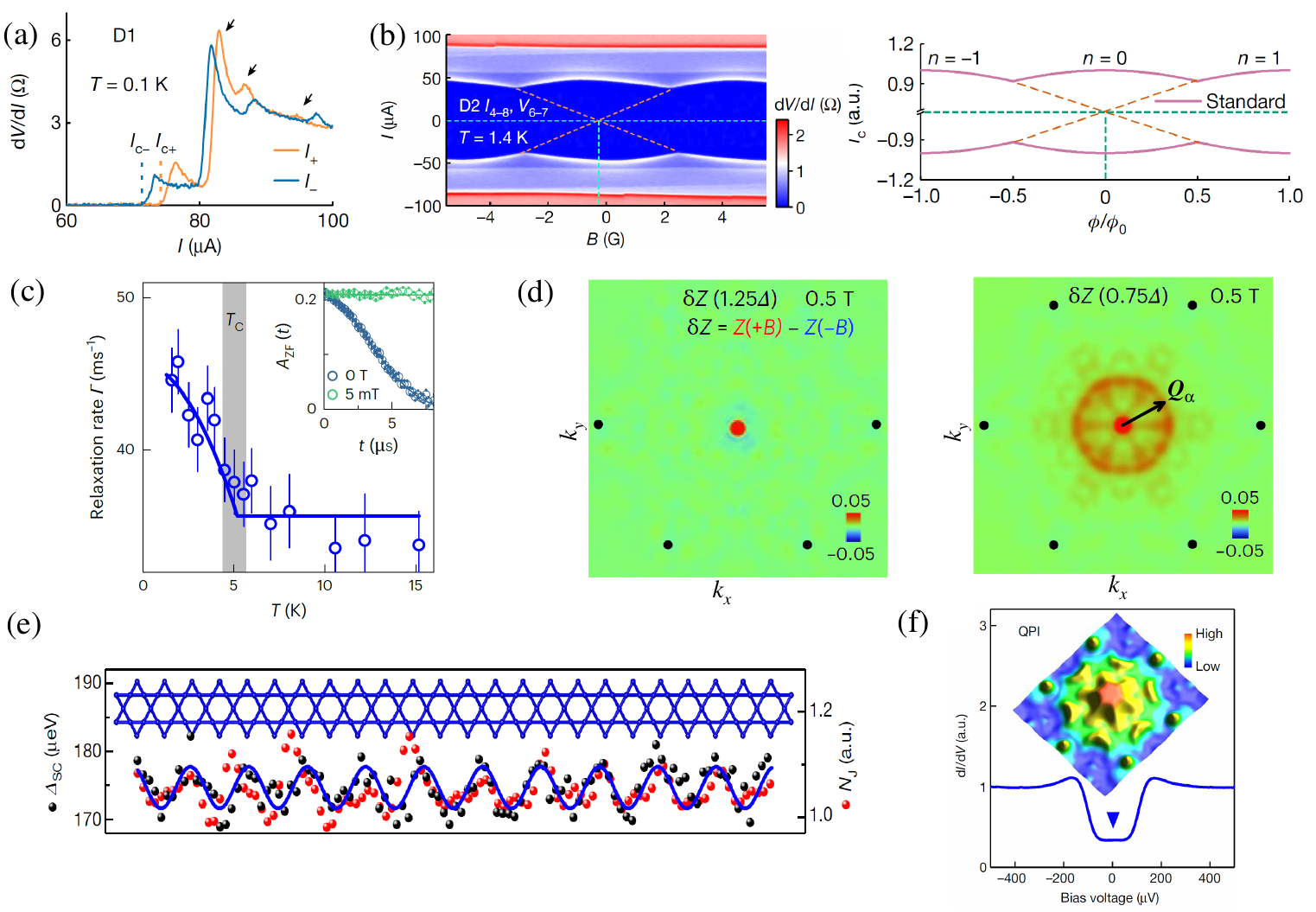}
    \caption{(a) Differential resistance ($dV/dI$) as a function of the d.c. current ($I$) at $0.1$ K. Adapted from Ref.~\cite{Nature.630.64}. (b) Left: Superconductivity interference pattern on a CsV$_3$Sb$_5$ flake measured at $1.4$ K. Right: Standard superconductivity interference pattern on a Little–Parks device. Adapted from Ref.~\cite{Nature.630.64}. (c) Temperature dependence of the zero-field muon spin relaxation rate for Ta-doped CsV$_3$Sb$_5$. The inset shows the $\mu$SR time spectra $A_{ZF}(t)$ measured in zero field and under a small magnetic field of 5 mT applied in a direction longitudinal to the muon spin polarization. Adapted from Ref.~\cite{Nat.Mat.23.1639}. (d) Time-reversal asymmetrical Bogoliubov quasi-particles interference signal $\Delta Z$ for $E=1.25\Delta$ (left) and $E=0.75\Delta$ (right). Adapted from Ref.~\cite{Nat.Mat.23.1639}. (e) Modulations of pairing-gap (black dots) and pair-density (red dots) with $2a$ periodicity. The inset is the kagome lattice reference. Adapted from Ref.~\cite{Nature.632.775}. (f) Evidence for residual Fermi arcs from in-gap differential conductance states and arc-like quasiparticle interference data at zero energy. Adapted from Ref.~\cite{Nature.632.775}.}
    \label{sc-fig-3}
\end{figure}

An intriguing unconventional property of the SC state in AV$_3$Sb$_5$ is its potential to break time-reversal symmetry, indicating it may be a chiral superconductor. Increasing experimental evidence supports this time-reversal symmetry breaking. For instance, the muon spin relaxation ($\mu$SR) experiments have shown that the time-reversal symmetry breaking feature of the CDW state in CsV$_3$Sb$_5$ persists into the SC state~\cite{Nature.602.245}. Additionally, transport measurements under a magnetic field reveal that CsV$_3$Sb$_5$ flakes exhibit chiral transport characteristics~\cite{npj.quantum.Mat.7.105}. Below, we will focus on recent experimental research progress in this area.

A recent study by Le \textit{et al}. demonstrates that CsV$_3$Sb$_5$ can exhibit the magnetic field-free SC diode effect~\cite{Nature.630.64}, characterized by nonreciprocal SC transport. Specifically, the critical current ($I_{c}$) varies depending on the direction of current flow. This effect is intrinsically linked to the time-reversal symmetry breaking, as it requires an asymmetry in the material's response to current flow. As shown in Fig.~\ref{sc-fig-3}(a), the critical current in the positive direction ($I_{c+}$) is larger than that in the negative direction ($I_{c-}$), indicating nonreciprocity. Moreover, the polarity of the zero-field SC diode effect, defined as $(I_{c+}-I_{c-})/(I_{c+}+I_{c-})$, can be altered by thermal cycling~\cite{Nature.630.64}. This suggests the presence of dynamic orders within an internal time-reversal symmetry breaking background. Since charge order is typically static near $T_c$, this finding is difficult to explain by chiral charge order inherited from the normal state. A plausible candidate is dynamic SC domains with time-reversal symmetry breaking. Consequently, these experimental results indicate that this SC state spontaneously breaks time-reversal symmetry, suggesting it is a chiral superconductivity. Furthermore, they found that the critical currents exhibit periodic oscillations under external magnetic fields, as shown in Fig.~\ref{sc-fig-3}(b). This phenomenon resembles the Little-Parks effect in a SC ring, where the critical current is modulated by a magnetic field with a period of a flux quantum $h/2e$~\cite{PhysRevLett.9.9}. This suggests the possibility of flux quantization within the CsV$_3$Sb$_5$ superconductor, indicating a SC domain structure with boundaries guiding the flow~\cite{Nature.630.64}. As shown in Fig.~\ref{sc-fig-3}(b), in contrast to the standard model, the pattern's symmetry center deviates from the zero field, further indicating time-reversal symmetry breaking in the SC state of CsV$_3$Sb$_5$.

The SC diode effect in CsV$_3$Sb$_5$ introduced above cannot entirely rule out the influence of chiral CDW order. Therefore, it is essential to confirm whether the superconductivity itself breaks time-reversal symmetry in samples without CDW order. In a recent study, Deng \textit{et al}~\cite{Nat.Mat.23.1639} utilized STM and $\mu$SR to demonstrate the time-reversal symmetry breaking superconductivity in Cs(V,Ta)$_3$Sb$_5$, where the CDW order is absent due to the partial replacement of V with Ta. As illustrated in Fig.~\ref{sc-fig-3}(c),  the zero-field muon spin relaxation rate significantly increases as temperature drops below $T_c$, indicating the emergence of a spontaneous magnetic field below $T_c$~\cite{Nat.Mat.23.1639}. This finding suggests that a time-reversal symmetry is broken in the SC state. A similar enhancement of the electronic relaxation rate below $T_c$ was also been observed in AV$_3$Sb$_5$ when the charge order is suppressed by pressure~\cite{Commun.Phys.5.232,Nat.Commun.14.153}. Additionally, STM measurements were performed to detect the time-reversal symmetry-breaking signal from the electronic structure. From the tunneling conductance $g(\bm{r},E)$, one can calculate the ratio $Z(\bm{r},E)=g(\bm{r},E)/g(\bm{r},-E)$ and its Fourier transformation $Z(\bm{q},E)$, which can enhance the Bogoliubov quasi-particle interference pattern~\cite{Nat.phys.3.865}. Figure~\ref{sc-fig-3}(d) shows the signal difference $\delta Z(\bm{q})=Z(\bm{q},+B)-Z(\bm{q},-B)$ between $Z(\bm{q})$ data with opposite $c$-axis magnetic fields $+B$ and $-B$~\cite{Nat.Mat.23.1639}. This quantity can characterize the time-reversal asymmetrical interference of Bogoliubov quasiparticles. For $E=0.75\Delta$, within the pairing gap, a clear time-reversal symmetry-breaking signal $\delta Z(\bm{q})$ at a circular vector $\bm{q}=\bm{Q}_{\alpha}$ was observed under $B=\pm0.5T$. In contrast, this signal disappears at energies outside the pairing gap ($E=1.25\Delta$). Thus, both $\mu$SR and STM experiments support the time-reversal symmetry breaking of the SC state in AV$_3$Sb$_5$ without CDW orders. The time-reversal symmetry breaking and the full SC gap are consistent with previous theoretical predictions of $d+id$ wave superconductivity~\cite{PhysRevB.85.144402,PhysRevB.87.115135,PhysRevB.86.121105,PhysRevLett.110.126405}.

In addition to the breaking of time reversal symmetry, Deng \textit{et al}. also reported an observation of SC order modulations in KV$_3$Sb$_5$ and CsV$_3$Sb$_5$ using STM~\cite{Nature.632.775}, offering evidence for the potential existence of pairing density waves in AV$_3$Sb$_5$. As illustrated in Fig.~\ref{sc-fig-3}(e), both the pairing gap and pairing density exhibit $2\times2$ modulations at the bulk charge-ordering vector of the kagome lattice. Furthermore, the intensities of the $2\times2$ vector peaks defines a clockwise chirality when counting from low to high, and the chirality of this pairing density wave can be switched by the magnetic field. These experimental findings indicate a chiral pairing-density-wave order with broken time-reversal symmetry. In contrast to conventional superconductors, the STM spectrum exhibits evident flat residual states inside the SC gap [see Fig.~\ref{sc-fig-3}(f)], which is also distinct from typical nodal superconductors, such as $d$-wave cuprate superconductors. By measuring quasiparticle interference, the researchers mapped the Bogoliubov Fermi surface, revealing arc-like structures consistent with the flat bottom in the density of states. The exotic pairing-density-wave state is proposed to arise from finite-momentum pairings between the Sb $p$ orbitals and V $d$ orbitals, induced by the couplings associated with the $2\times2$ CDW order~\cite{Nature.632.775}.

\section{Summary and perspective}\label{sec-su}

In this paper, we have reviewed recent advancements in the understanding of QSLs, fractional magnetization plateau phases in kagome antiferromagnets, and unconventional superconductivity in the kagome materials AV$_3$Sb$_5$. Recent theoretical studies have increasingly suggested that the ground state of the NN kagome AF Heisenberg model is a gapless Dirac QSL. Moreover, when more complex interactions are introduced, more exotic spin liquids, such as chiral spin liquids, may also emerge in kagome antiferromagnets. Under an external magnetic field, kagome antiferromagnets can exhibit multiple fractional magnetization plateaus, with the 1/9 magnetization plateau phase potentially being a $\mathbb{Z}_{3}$ QSL. In terms of kagome superconductivity, both weak-coupling approaches, which focus on the electronic structure, and strong-coupling approaches, which emphasize local correlations and doped spin liquids, suggest that superconductivity in the kagome system is unconventional. For vanadium-based kagome superconductors AV$_3$Sb$_5$, accumulating experimental evidence indicates a rich landscape of unconventional superconductivity, with recent experiments providing evidence of chiral superconductivity and pairing density wave states.

Despite this exciting progress, significant challenges remain. For example, while materials like YCu$_3$(OH)$_6$Cl$_3$ and YCu$_3$(OH)$_6$Br$_3$ possess nearly ideal Cu$^{2+}$ kagome planes due to the distinct ionic radii of  Y$^{3+}$ and Cu$^{2+}$, the kagome layers still experience considerable bond disorder from the mixed occupancy of Cl$^{-}$, Br$^{-}$ and (OH)$^{-}$  anions. Thus, a systematic investigation of disorder effects is essential for a comprehensive understanding of kagome antiferromagnets, particularly through theoretical research. In addition, the momenta of Dirac cones shown in the inelastic neutron scattering spectra of the kagome magnetic material YCu$_3$(OH)$_2$Br$_2$[Br$_{1-x}$(OH)$_x$] differ significantly from those predicted for the Dirac QSL in the NN kagome AF Heisenberg model~\cite{nat.phys.20.1097}. The spectra further indicate that the Dirac excitations are associated with a tripled unit cell, resembling that of the $\mathbb{Z}_{3}$ QSL, and that the Dirac cones are sharp. These features suggest that the Dirac QSL realized in this material may involve a previously unexplored mechanism, possibly associated with an enlarged unit cell or additional interactions beyond simple model, highlighting an intriguing direction for future theoretical and experimental studies.

The chiral superconductivity and pair density wave states observed in AV$_3$Sb$_5$ call for further experimental evidences and investigations to clarify their origins and properties. Key unresolved issues include the precise structure of the superconducting gap function and the respective roles of electron–phonon coupling versus electron correlations in the pairing mechanism. Moreover, superconductivity in AV$_3$Sb$_5$ is intertwined with time-reversal–symmetry–breaking charge density waves and electronic nematicity, yet the relationships between superconductivity and these intertwined orders remain elusive. Existing theoretical studies primarily focuse on the V $3d$ orbitals, often based on single- or two-orbital models; however, the Sb $5p$ orbitals also make significant contributions to the Fermi surface. Incorporating the Sb $5p$ degrees of freedom together with the V $3d$ orbitals may therefore be important for understanding many of the unconventional properties of AV$_3$Sb$_5$. In particular, while the spin–orbit coupling of the V $3d$ orbitals is relatively weak, the much stronger spin–orbit coupling associated with the Sb $5p$ orbitals could play a noticeable role in the emergence of these unconventional phenomena.

We anticipate that exploring these avenues will uncover novel physics in kagome quantum spin liquids and superconductors. These investigations are expected to enhance our understanding of strongly correlated electron systems and may inform the design of new quantum materials with unique topological and superconducting properties.

\backmatter

\bmhead{Acknowledgements}

We thank Z.-X. Liu for the helpful discussions.

\bmhead{Author contributions}

LWH and SLY drafted the first version of manuscript. JXL supervised this project. All authors read and approved the final manuscript.

\bmhead{Funding information}

This work was supported by National Key Projects for Research and Development of China (No. 2021YFA1400400 and No. 2024YFA1408104) and the National Natural Science Foundation of China (No. 12374137, No. 12434005, No. 92165205, and No. 12074175).

\bmhead{Availability of data and materials}

All materials used in this paper are available on reasonable request.

\section*{Declarations}

\begin{itemize}
\item Competing interests

All authors declare that there are no competing interests.

\end{itemize}

\bibliography{kagome-bib}

\end{document}